**University of Kirkuk**

**College of Agriculture**

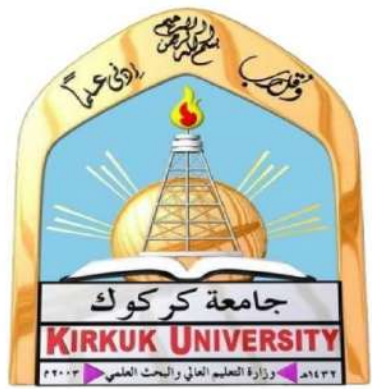

# EFFECT OF ORGANIC FERTILIZER AND NANO CALCIUM SPRAY ON GROWTH, YIELD AND STORAGE CHARACTERISTICS OF FIG FRUITS CV. WAZIRI

A Thesis Submitted by

**Aram Sabir Ahmed Fatah**

B. Sc. in plant production – Sulaimani Polytechnic University (2011-2012)

To

The Council of The College of Agriculture, University of Kirkuk

As Partial Fulfillment of the Requirements

For The Master Degree in Agricultural Sciences

**(Horticulture & Landscape Design)**

**Supervised by**

**Dr. Mohamad Abdul Aziz Lateef**

**Assistant professor**

| | |
|---|---|
| 2025 A.D. | 1447 A.H. |

بِسْمِ اللَّهِ الرَّحْمَٰنِ الرَّحِيمِ

﴿ وَالتِّينِ وَالزَّيْتُونِ ۞ وَطُورِ سِينِينَ ۞ وَهَٰذَا الْبَلَدِ الْأَمِينِ ۞ لَقَدْ خَلَقْنَا الْإِنسَانَ فِي أَحْسَنِ تَقْوِيمٍ ۞ ثُمَّ رَدَدْنَاهُ أَسْفَلَ سَافِلِينَ ۞ إِلَّا الَّذِينَ آمَنُوا وَعَمِلُوا الصَّالِحَاتِ فَلَهُمْ أَجْرٌ غَيْرُ مَمْنُونٍ ۞ فَمَا يُكَذِّبُكَ بَعْدُ بِالدِّينِ ۞ أَلَيْسَ اللَّهُ بِأَحْكَمِ الْحَاكِمِينَ ۞ ﴾

صدق الله العظيم

**سورة التين الآية (١-٨)**

# **Dedication**

To everyone who taught me a letter... my esteemed teachers.

To the meaning of love, compassion, and devotion... to the smile of life and the secret of existence—my dear father.  Allah grants him paradise.

To the one who supported me with her prayers, who stayed up countless nights to light my path, the source of kindness and affection, my beloved mother. May Allah bless her with a long life.

To the one who was my greatest support throughout my academic and research journey, my lifelong companion, and the light of my eyes, who endured so much with me and for me—my beloved wife (Bave).

To those who adorned my life and are my greatest legacy, the most beautiful gifts from my Creator—my son, Avyar, and my daughter, Yavan.

To my dear brothers (Omed, Hiwa, Nabaz, and Bahroz), whose presence made every challenge easier.

My dear sisters (Zakiyah, Suaad, Rizan, Niyaz, Shelan, and Shokhan) have always supported and encouraged me to keep going.

To those who stood by me on this journey, contributed to my education, and provided my constant support, I would like to express my gratitude to my esteemed supervisor, Dr. Mohammed Abdulaziz Lateef.

My dear parents-in-law have always been there to support me through life's most challenging moments. May Allah bless them with a long life.

To everyone who waited for my success and rejoiced, to everyone who helped me in any way, and even to those who caused me pain and sorrow—thank you, and you have my deepest respect.

# Acknowledgment and Appreciation

All praise is due to Allah, the Lord of all worlds, and may peace and blessings be upon our Prophet Muhammad, his pure family, his noble companions, and those who follow his guidance until the Day of Judgment.

I extend my sincere gratitude to the University of Kirkuk, College of Agriculture, for allowing me to pursue my master's degree. My deepest appreciation goes to my esteemed supervisor, Asst. Prof. Dr. Mohammed Abdulaziz Lateef, for his invaluable guidance, tireless support, and insightful advice throughout my research journey. His encouragement and scientific expertise were instrumental in overcoming challenges, and for that, I express my utmost gratitude and respect. May Allah reward him abundantly.

I also extend my sincere appreciation to the Department of Horticulture and Landscape Engineering for their continuous support throughout my studies, particularly Prof. Dr. Kefaia Ghazi Alsaad. My gratitude goes to the Assistant. Prof. Dr. Ali Mohammed Noori, head of the Plant Tissue Culture Laboratory at the College of Agriculture, for his assistance in sample analysis and for providing all available resources to complete the necessary laboratory tests.

I am also sincerely grateful to the Chairman and members of my thesis defense committee for agreeing to review and evaluate my work. Their insightful comments, scientific observations, and constructive feedback have significantly enriched my research and contributed to its improvement.

My heartfelt appreciation and love go to my beloved family and my in-laws for their unwavering emotional and financial support, as well as their unmatched assistance throughout this journey.

# *Aram*

**Summary**

   This study was carried out to estimate the effect of organic fertilizer and Nano Calcium spray on the growth, yield, and storage of fig fruits cv. Waziri. It was conducted in the village of Kani Sard in the Sharbazhir district near the Sitak area. This is about 35 km northeast of the center of Sulaymaniyah (latitude: 35°38'25.7"N, longitude: 45°35'12.0"E, altitude). This study was conducted at 54 plants with three-year-old fig trees cultivated using the cordon method during the 2024 growth season. The controlled environment of the plastic house enabled a precise evaluation of how different fertilizer treatments affected plant development, fruit traits, and post-storage fruit quality. A randomized complete block design (R.C.B.D.) factorial laid out within two main factors: Bio Health fertilizer applied at three levels (0, 15, and 30 g tree$^{-1}$) and Nano Calcium applied at three concentrations (0, 75, and 150 mg L$^{-1}$). The Bio Health fertilizer was applied to the plant in the soil, and Nano Calcium as the foliage with a 16-liter backpack sprayer until completely wet. The fertilizer was supplied in two batches during the growing season, with sprayed in three batches, including the first added (16/4/2024) and the second on (16/5/2024), and the first spraying after a week of fruit holding (25/5/2024). The second spray was after 15 days of the first one, and the third after 15 days of the second spraying. The study reached the following results:

1. Bio Health (15 g tree$^{-1}$): soil application of 15g tree$^{-1}$ Bio Health led to superiority in some characteristics, such as fruit hardness before storage, recorded the highest value of 8.612N. On the other hand, the highest value of SSC was 945.194 µg Glu.g$^{-1}$ FW. Furthermore, the highest amount of AC was 224.600 µg TE.g$^{-1}$ FW. For carotenoids, the highest value was 2.495 µg g$^{-1}$ FW recorded. After storage, the highest TSS amount was 18.372°Brix

2. Bio Health (30 g tree$^{-1}$): application of soil of 30g tree$^{-1}$ Bio Health, especially meaningfully improved most vegetative growth characteristics (the highest value of leaf area was 538.82 cm$^2$, for leaf chlorophyll content recorded the highest value was 22.52 mg g$^{-1}$ fresh leaf weight, leaf dry weight was 33.05g, and leaf nitrogen was recorded with the highest value of content at 1.879%. The highest amount of phosphorus was 0.393% in the leaf. The highest value of potassium was 1.052%). The fruit quantitative and characteristics (fruit weight, the highest value was 104.76g. The results of fruit volume were 115.03 cm³, whereas the smallest volume, 88.57 cm³. Yield tree produced the highest yield, 10.436 kg tree$^{-1}$) Fig phytochemical characteristics (Before fruit storage, the results confirmed that the highest value of TSS recorded was 18.33 °Brix. However, after storage of fruit, the highest amount of SSC, AC, ASA, carotenoid, TCC, and TA were 690.740 µg Glu.g$^{-1}$ FW, 598.19 µg TE. g$^{-1}$ FW, 29.830 mg g$^{-1}$ FW, 1.864 µg g$^{-1}$ FW, 28.637 g Glu 100 g$^{-1}$ FW, and 0.522% citric acid).

# SUMMARY

3. Nano Calcium (75 mg L$^{-1}$): resulted in supremacy in several traits, including for (AC and Carotid) after the use of (75mg L$^{-1}$), which were (246.106 µg TE. g$^{-1}$ FW and 2.514 µg g$^{-1}$ FW). Furthermore, after fruit storage, the highest value of AC was 582.57 µg QE. g$^{-1}$ FW, for SSC, the highest value was 706.862µg Glu.g$^{-1}$ FW).

4. Nano Calcium (150 mg L$^{-1}$): Nano Calcium foliar spray significantly affected vegetative growth characteristics. For leaf area, the highest value was 507.73 cm$^2$. Total chlorophyll content was recorded as 21.13 mg g$^{-1}$ fresh leaf weight. For leaf dry weight, the highest value was 31.25g, which significantly proved that the highest nitrogen value was 1.678%. The highest value of phosphorus was 0.366%. The highest amount of potassium was 1.004%. The fruit's quantitative characteristics (the maximum value of fruit weight of 96.55g. The highest fruit volume was 107.46 cm³. The highest tree yield was 8.881 kg tree$^{-1}$, compared to the control. The maximum amount of fruit moisture was 82.052% at the use of 150mg). L$^{-1}$ Nano Calcium. After the fruit was stored, it showed the highest value of hardness was 9.323N). Fig phytochemical characteristics before storage results showed a significant impact on (ASA, TCC, and TSS) parameters, which had high rates (1.905 mg g$^{-1}$FW, 24.695g 100 g$^{-1}$ FW, and 17.45°Brix). TCC, after storage, was recorded as the highest value, 26.240g 100 g$^{-1}$ FW).

5. The interaction between Bio Health 30 g tree$^{-1}$ and Nano Calcium 150 mg L$^{-1}$ significantly increased vegetative growth characteristics (leaf area had the highest value of 566.15 cm$^2$, the highest amount of chlorophyll in the leaf was 24.68 mg g$^{-1}$ fresh leaf weight. However, leaf dry weight was the highest at 37.59g. The highest value of leaf nitrogen content, which was 1.947%, was observed, showing that the highest amount of phosphorus was 0.399%. For the amount of leaf potassium, the results also recorded the highest amount of leaf potassium at 1.150%). The fruit's quantitative characteristics (the highest value of fruit weight was 110.13g, at the same time, the most significant fruit volume of 120.00 cm³ was verified. The highest value of yield was 11.113 kg tree$^{-1}$, and the maximum amount of fruit moisture at 82.543%. However, the largest fig fruit size was about 57mm$^3$). Fig phytochemical characteristics' results recorded the highest values of total phenol content and antioxidant capacity, 199.750 µg GAE.g$^{-1}$ and 678.453 µg TE.g$^{-1}$ FW after fruit storage. Total soluble solids (TSS) before storage, the highest value was 19.23 °Brix. The highest value of total carbohydrate content before storage (TCC) was 26.107 g Glu. 100 g$^{-1}$ FW. However, for TCC after fruit storage, the highest value was 30.129 g 100 g$^{-1}$ FW.

# LIST OF CONTENTS

## LEST OF CONTENTS

## LEST OF CONTENT

# LEST OF CONTENTS

# LIST OF TABLES AND FIGURES

## List of Tables

**LEST OF TABLES**

**LIST OF FIGURES**

## 1.    Introduction

The common fig (*Ficus carica* L.) is a deciduous tree belonging to the Moraceae family. It is one of the first fruit plants to be domesticated and an essential crop worldwide for fresh and dry consumption (Kamiloglu and Capanoglu, 2015). Fig fruit is a highly nutritious food for human health due to its high nutrient value, fiber content, and laxative properties; Therefore, it is consumed as fresh or dry worldwide, even though figs can easily adapt to climatic conditions, their growth and development change depending on the environment (Ammar *et al.*, 2020). The high concentration of carbohydrates, vitamins, and minerals in fig fruits makes them extremely nutritious; the majority of fig-growing regions don't have any irrigation or fertilization schedule that meets the water and nutrient needs for optimal tree growth, high yields, and improved fruit quality (Mustafa *et al.*, 2022).  It originated in South Asia, with the highest possibility, and then gradually developed into the Mediterranean region, Central Asia, and Transcaucasia (Mars, 2003; Mohammed, 2025). Figs are one of the five fruit plants mentioned in the Holy Quran (Sheikh, 2016), Researchers have also confirmed that figs are considered among the most important edible horticultural fruits in terms of economic, ecological, and industrial studies, and are abundant in nutrients vital for human health. According to FAOSTAT (2024), the total fig production around the world was 1,242,449 tonnes in 2022. The figs have a variety of skin colors, ranging from green to deep purple. Although figs can be eaten whole, they are usually peeled, with the pulp being consumed and the skin discarded (Hussaini *et al.,* 2020).  Fruit can be stored for up to two or four weeks by combining a cooler and a $CO_2$ enriched atmosphere, Fresh fruits have a limited post-harvest life of seven to ten days by nature, due to their longer shelf life after drying, figs are also quite popular as dried fruit (Solomon *et al.,* 2006; Kamiloglu and Capanoglu, 2015). In addition, it is the first plant cultivated by humans, fig is also used raw, dried, canned, or in other preserved forms, Morphologically, figs comprise both trees and shrubs, The bark is smooth and grey, the common shape of fig is turbinate obovoid, the colors are green-yellow, copper, creamy, red, and florid, depending on genotypes and variety, In the last decade, researchers focused on numerous recent tools, to progress in sustainable horticultural fruit production, including bio stimulants and Nano-fertilizers, This can significantly improve fruit quality parameters and make nutrients more available (Colla *et al.,* 2015; Rouphael and Colla, 2020). Therefore, the use of BOMFs Bio-Organic Mineral Fertilizer is becoming increasingly significant in modern agriculture, these fertilizers aim to make nutrient use efficient for the plant by decreasing the quantity of artificial fertilizer used with cultivation costs, subsequently; they increase nutrient use efficiency, which has safe and eco-friendly ecosystems and human health (Ayenew *et al.,* 2024). As mentioned before, one of the tech innovations is bio stimulant, to achieve a sustainable increase in horticulture food production, which contains natural-origin compounds or microbes, to stimulate plant processes to improve nutrient use efficiency and tolerance to abiotic and biotic stresses (Drobek *et al.,* 2019). Bio Health WSG is a water-soluble, organic fertilizer that is used as a good bio stimulant, it is based on humic acid, seaweed, and microorganisms (beneficial bacteria and fungi), This combination can generally affect the properties of the chemo-physical activity of soil and the increased efficiency rhizosphere to availability of more nutrients by improvement of root growth and morphology, physiology, and intolerance against abiotic stressors (Vujinovi *et al.,* 2020). Furthermore, beneficial bacteria and fungi can have many important impacts, including developing hormonal balance within plants and the biosynthesis of volatile organic compounds (Shahrajabian *et al.*, 2021). A progressed system of tolerance to abiotic stresses by induced auxins and secondary metabolites (Ruzzi and Aroca, 2015, and Fiorentino *et al.,* 2018).

Nano fertilizers are another new stage that is applied in very low doses with a high absorption rate compared to other fertilizers without a negative impact on plant development with growth, nutritional status, and the ecosystem (Zhang, *et al.*, 2019a; Bian, *et al.*, 2020; Arora, *et al.*, 2018). The 21st century saw the development of nanomaterials, which come in several forms and quantities, including synthetic and natural materials that are either organic or non-organic, Applications of nanomaterials technology in agriculture, particularly on fruit trees, could secure the sustainability of both the environment and agriculture, (Al-Hchami and Alrawi, 2020).

Therefore, one of the important nutrients and crucial roles in the development of growth fig trees is calcium ions ($Ca^{+2}$), This has a major impact on the improved fruit quality and nutrition both during harvest and storage, It is also important to control physiological processes in plants, including the root hair lengthening, the formation of pollen tubes, and the movement of stomatal guard cells, cell walls, and membranes, which act as an intracellular messenger within the cell (Raza, *et al*.,2020; Jiang, *et al*.,2020; Shah, *et al*., 2019 and Huang, *et al.*, 2021).

Furthermore, in fig orchards, pre-harvest calcium spraying is a potential cultural practice; Cross-links between calcium and pectin help stabilize cell wall structures and prevent enzymes from breaking them down (Zhao, *et al*., 2019; Zhou, *et al*., 2020 and Jiang *et al*., 2022). Calcium plays a crucial role in the growth and development of fig trees and has a major impact on the fruit's quality It is improved by calcium nutrition both during harvest and storage, in fig orchards, pre-harvest calcium spraying is a potential cultural practice (Souza *et al.*, 2023).

Eventually, they found a good positive correlation between the applied rates of the Nano-Bio fertilizers and the tree's vegetative growth and productivity, Consumption of fruits and vegetables has improved recently as a result of consumers relating them to a lower risk of serious illnesses and a potential delay in the start of age-related disorders (Crisosto *et al*., 2010; Pereira *et al*., 2017). Figs are seasonal fruits belonging to the Moraceae family that can be harvested twice a year in spring or summer, The thin peel and fleshy fruit of the fig make it difficult to store and transport, it is susceptible to rotting and deterioration; Therefore, it is crucial to advance fig preservation and fresh fruit storage technology; Thus, the improvement and systemization of storage technology are significantly impacted by the study of pre-harvest processing technology to enhance the quality of fig storage, (Zhang *et al*., 2019b; Kjellberg and Lesne, 2020).

Thus, this study was conducted to show the impact of Bio Health and Nano Calcium fertilizer on fig vegetative growth and fruit qualities and quantities, which can be summarized as follows.

1. To enhance the fig vegetative growth by adding the Bio Health fertilizer, and improve the fig fruit qualities by adding Calcium elements in the shape of Nano.

2. To prolong the shelf life by adding Calcium elements in the shape of Nano fertilizer.

3. To extend the storage life of fig fruits and prevent them from fast decay.

## 2. Literature Review

### 2.1 Fig Tree

The fig tree (*Ficus carica*) is classified as a deciduous, subtropical tree, often cultivated with multiple stems, and branches are typically non-intertwining, unlike many other tree species also fig trees are characterized by the presence of a milky latex, a distinctive sap with a notable odor that is secreted when a leaf or fruit is cut, and this latex is caustic and pungent, dries upon exposure to air, and serves as a natural source of rubber, the root system of fig trees is generally fibrous, allowing the tree to adapt to a range of soil types, however, it performs best in deep, well-drained loamy soils with moderate fertility, Soils with good water retention and adequate aeration, such as sandy loam or clay loam, are ideal; in contrast, highly saline, poorly drained, or heavy clay soils can inhibit root growth and reduce productivity and the leaves are relatively large, palmately shaped, and can be either entire or lobed, depending on the cultivar while typically, the leaves are three-lobed, and in the axil of each leaf, there are usually one to three buds; the central bud is generally vegetative, while the lateral buds are reproductive (Cook and Rasplus, 2003; Ronsted *et al.*, 2005). From an environmental perspective, fig trees thrive in warm, dry climates with long summers and mild winters, the optimal temperature range for their growth is 15°C to 40°C Although they are relatively drought-tolerant once established, they benefit from periodic irrigation during extended dry spells, especially during fruit development and full sunlight exposure at least 6 to 8 hours per day is essential to ensure proper growth and fruit quality, while Regarding reproduction, fig flowers are classified into four types: male, female, sterile, and gall (or caprifig) flowers, and the fig fruit is a false fruit, botanically referred to as a syconium, it is a fleshy receptacle that encloses a hollow cavity, connected to the outside by a small opening known as the ostiole (Machado *et al*., 2005; Haine *et al*., 2006).

### 2.2 Organic Fertilizers (Bio Health):

The modern agricultural approach relies on the use of organic fertilizers to increase production and reduce environmental pollution resulting from the excessive use of chemical fertilizers, as humus is a natural source rich in nutrients necessary for plant growth, which consists mainly of humic acid, fulvic acid, and humin (Ferrara and Brunetti, 2010). Despite the efficiency of chemical fertilizers in increasing production and improving quality, it has recently been proven that they have harmful effects on human health, and the modern trend is to reduce the use of chemical fertilizers and add organic compounds that are harmless to the environment and human health and increase the resistance of plants to harsh environmental conditions (Shehata *et al*., 2011). Among the modern techniques adopted to improve soil fertility and crop quality is the role of organic fertilizers such as Bio Health; these organic fertilizers are characterized by containing natural components that enhance the biological activity of the soil and increase the efficiency of nutrient absorption, which improves plant growth and yield (Akinci and Simsek, 2019).

Bio Health is a water-soluble organic fertilizer used as a good bio stimulant. Based on humic acid, seaweed, and microorganisms (beneficial bacteria and fungi), this combination can generally affect the physical and chemical activity properties of the soil and increase the efficiency of the roots to provide more nutrients by improving growth (Vujinovic *et al*., 2020). It has been found through studies that it improves soil fertility and thus increases nutrient availability, which increases plant growth and yield (Hartwigson and Evans, 2000). Organic humic substances have an effective effect in improving the chemical and physical properties of the soil by interacting these substances with soil minerals (Mataroiev, 2002), in addition, these organic humic acids have a chelating role and work to increase the availability of mineral elements in the soil and thus increase plant growth.

Organic humic fertilizers contain, in their various sources, a large and wide range of organic compounds that are soluble in water, such as sugars, amino acids, humic and non-humic organic acids, and proteins; all of these compounds contribute directly or indirectly to the growth and development of the plant they are either in the form of growth stimulants due to the enzymatic or hormonal effect, this is because they contain the nutrients that the plant needs, or they have an effect in increasing the availability of these nutrients that are present in the soil or added to it, which leads to improving its quality and increasing production (Fartusi, 2003).

Many studies have also shown that humic substances, Fulvic acid, and Humic acid are the main components (70-65%) of organic matter (Yildirim, 2007). Organic fertilizers are natural materials that can encourage vegetative and root growth and improve the nutritional status of plants (Vernieri *et al.,* 2006). It improves the physical, chemical, and biological properties of the soil and its ability to retain water, and increases the availability of nutrients, as it acts as a storehouse for these elements in addition to reducing the pH of the soil, which leads to facilitating these elements in the soil (Agbede *et al*., 2008).

Organic fertilizers have a wide impact on plant growth through their effect on the processes of respiration and photosynthesis (Dantas *et al*., 2007). Marine extracts contain many nutrients and growth regulators that stimulate growth at low concentrations and do not leave any harmful residues for humans, animals, or the environment (Spinelli *et al*., 2009). Organic fertilizers are widely used in agriculture, and most published and reliable research has proven that the effect of humic is similar to the impact of hormones, especially auxins, cytokinins, and gibberellins (Serenella *et al*., 2009). The composition of the marine extract depends on the raw material, the geographical location from which it is taken, the type of algae, the extraction method, and the number of applications (Al-Musawi, 2018).

### 2.2.1 Influence of Organic Fertilizer on vegetative characteristics.

Thran and Kose, (2004) indicated that grape vines with the addition of seaweed extracts, Maxi crop, and Alga Powder at concentrations of (0, 0.5, 1, and 2) ml $L^{-1}$ led to an increase in the concentrations of some significant elements such as nitrogen, phosphorus, and potassium.

Venkata *et al*., (2009) found that white mulberry trees, variety V1, by using organic extract vermi wash led to improved growth characteristics and significantly increased leaf area, total chlorophyll, and mineral content N, P, and K in leaves.

Osman and Abd El-Rhman, (2010) recorded in this study that the maximum area of the fig leaf was 540 cm³ when 400 g of poultry manure and ozospirillum were added to the plants, while the minimum leaf area was 390 cm³ at the control.

According to Fathy and El-Shall, (2010), 8-year-old apricot trees of the Conino variety were treated with Actosol 2.9% humic acid, a soil addition treatment at a concentration of 9.0 mL L-1, with a one-week difference during the growing season. This led to a significant increase in leaf area, respectively, compared to the control treatment, which gave the lowest value.

Al-Mousawi (2011), obtained a significant increase in the content of leaves of nitrogen and phosphorus when treating Turkish fig seedlings by adding agroleaf nutrient solution at a concentration of (15) g $L^{-1}$.

As reported by Abd EL-Razek *et al*., (2012), the application of humic acid on peach trees at two levels (0.25% and 0.50%) showed that the higher concentration (0.50%) significantly increased leaf chlorophyll content to, as well as the mineral content of nitrogen, phosphorus, and potassium.

Zhang *et al*., (2013) found in a study on Fuji Red apple trees that treating with humic acid at concentrations of 50, 100, 150, 200, 250, and 300 mg $kg^{-1}$ significantly increased the dry weight percentage of leaves, particularly at 200 mg kg$^{-1}$. This concentration also enhanced leaf chlorophyll content and the mineral content of nitrogen (N), phosphorus (P), and potassium (K).

Parkash *et al*., (2013) confirmed in a study that included the effect of several concentrations of Potassium humate on white mulberry trees and concluded that the addition of 4% concentration achieved a significant increase in leaf area, dry weight of leaves, mineral content N, P, K and total chlorophyll in leaves.

Kar *et al*., (2014) confirmed that the addition of humic acid at a concentration of 0.01% (Metallic humate) to white mulberry variety S-1635 significantly increased the plant's content of mineral elements P and K.

Maria and Evangeline, (2015) found a study showing the effect of marine extract SLF on some characteristics of white mulberry (*Morus alba* L.) and the results of the study showed that marine extract SLF at a concentration of (1%) led to a significant increase in leaf area and total chlorophyll content.

Ram *et al*., (2017) indicated in a study conducted to know the effect of potassium humate and farmyard manure on some growth characteristics of white mulberry trees, variety Bc259, that adding (10 kg hectare$^{-1}$) of KH significantly increased branch length, leaf area, dry weight, total chlorophyll content, and mineral content of N, P, and K in leaves.

The study of Hadi and Khalil, (2017) on the effect of adding organic fertilization on increasing the chlorophyll content of the leaves of the Halwani grape variety (*Vitis vinifera* L.), which was reached when compared with the control. An increase in the percentage of nitrogen in the leaves of the Halwani grape variety when compared with the comparison treatment, and an increase in the leaf area of the Halwani grape variety when compared with the comparison treatment. An increase in the percentage of phosphorus in the leaves of the Halwani grape variety when compared with the comparison treatment, and an increase in the percentage of potassium in the leaves of the Halwani grape variety when compared with the control treatment.

Ram and Maji (2018) found that adding 25 kg ha$^{-1}$ of Potassium humate improved the growth characteristics of white mulberry trees, variety Bc259, and significantly increased leaf area, total chlorophyll, and mineral content of the elements N, P, and K in the leaves.

The research (Aksoy *et al*., 2022). The vegetative growth parameters of fig trees were considerably enhanced by treatment with organic fertilizers. When organic fertilizer was applied, the amount of chlorophyll in the leaves grew significantly, reaching values of up to 63.10 mg per 100 g of fresh leaf weight. When coupled with humic acid, this improvement was even more pronounced, as the chlorophyll concentration increased to 74.00 mg/100 g, indicating increased photosynthetic efficiency. Increased levels of potassium (3.39%), phosphorus (0.45%), and nitrogen (2.08%) were found in the leaf nutrient analysis. These findings demonstrate that organic fertilization improves fig trees' overall development performance by having a positive impact on their morphological and physiological characteristics.

El-Sayed and Farouk, (2023) According to this study, using organic fertilizer greatly improved the fig trees' vegetative growth; for example, their leaf area increased from 27.4 to 39.1 cm², and their chlorophyll content increased in SPAD units. Additionally, their dry leaf weight improved, and nutrient analysis showed that their leaves had higher levels of nitrogen, phosphorus, and potassium than those of the controls, which led to healthier plants and better

Jafari *et al*., (2024) The effect of organic fertilizers significantly enhanced the vegetative growth of fig trees, leaf content of nitrogen, phosphorus, and potassium varied from 0.69 % in added sheep manure to 0.88% in chicken, 0.044% in the control, to 0.124 % in the vermicompost, and 31.6g Kg$^{-1}$ DW in the cow manure to 36.6g Kg$^{-1}$ DW in the chicken, respectively.

## 2.2.2 Effect of Organic Fertilizer on fruit qualitative and quantitative characteristics:

Fathy and El-Shall, (2010) showed that Canino apricot trees treated with a concentration of 1.5 ml $L^{-1}$ with Actosol 2.9% humic acid weekly during the growth stage led to a decrease in the percentage of total acidity, compared to the control treatment. It led to a significant increase in the percentage of total soluble solids (TSS) in the fruits and concluded in their study that when effect of Actosol humic acid 2.9 at a concentration of 15 ml $L^{-1}$ and repeating it weekly during the growing season on apricot trees of the Canino variety, led to a significant increase in the tree yield and the weight of a single fruit.

Abu Nuqta and Batha, (2010) concluded in the study they conducted on Halwani grape vines that fertilization with potassium humate at a concentration of 1 g $L^{-1}$ led to a significant increase in the percentage of total soluble solids (TSS). It did not affect the percentage of total acidity and its stability, and led to a decrease in the percentage of vitamin C compared to the comparison treatment when spraying with potassium humate on the vines four times.

Abd-El-Razek *et al*., (2012) indicated in their study on three-year-old Florida Prince peach trees for the two seasons (2011-2010) when using humic acid by adding it to the soil at two concentrations (0.25% and 0.50 %). A significant increase was obtained in the percentage of total dissolved solids. They noticed a significant decrease when using humic acid in the total acidity percentage of the fruits, compared to the control treatment, which recorded the highest percentage, respectively.

Ibrahim and Ali, (2016) studied the effect of humic acid on grape characteristics and yield using the commercial product (Humatic8500) (potassium humate, amino acid, Seaweed extract) and concluded that adding humic acid at a concentration of 3 g $L^{-1}$ three times significantly increased the weight and size of the fruit, the percentage of soluble solids, total sugars, and the total yield.

Berot *et al*., (2017) stated in his study on the effect of adding some organic fertilizers on the growth and fruiting of Royal apricot trees during the growing season (2013-2014) when using humic acid at concentrations of (0, 15, and 30) ml $L^{-1}$, it gave the highest rate in the number of fruits remaining on the trees at harvest, respectively, and gave the highest yield per tree, respectively.

Karim *et al*., (2019) evaluated the effect of humic substances on fig trees and found that adding humic acid at 3 g $L^1$ significantly increased the average fruit weight (from 31.2 g to 39.8 g) and fruit volume (from 26.5 cm³ to 33.4 cm³). Moisture content rose to 80.6%, while fruit firmness improved from 5.7 N to 7.9 N. The treatment also enhanced total yield per tree by 21.6% and reduced fruit drop by 17.4% compared to the control.

Hassan and Mohammed, (2020) investigated the effects of potassium humate, a substance that is comparable to Bio Health, on the yield and quality of fig fruit. A substantial increase in fruit weight (from 33.5 g to 42.6 g), fruit volume (from 28.4 cm³ to 35.7 cm³), and moisture content (from 76.2% to 81.4%) was seen when potassium humate was applied four times during the growing season at a rate of 2.5 g $L^1$. Additionally, fruit firmness improved from 6.1 N to 8.4 N, fruit drop decreased by 19%, and overall yield per tree increased by 23.8%.

Salem and Fathy, (2021) reported that fig trees treated with organic fertilizer rich in humic acid (Bio Grow) at a rate of 4 L $tree^{-1}$ showed significant increases in fruit weight (43.2 g), fruit volume (36.9 cm³), and total yield per tree (up to 18.7 kg). Moisture content was 82.1%, fruit firmness reached 8.7 N, and fruit drop decreased by 20.5% compared to untreated trees.

Najim *et al*., (2022) demonstrated that using potassium humate (2 g $L^1$) improved several fruit quality traits in fig trees: fruit weight increased from 30.5 g to 41.2 g, fruit firmness from 5.5 N to 7.8 N, and yield per tree rose by 24.3%. Moisture content reached 81.7%, and fruit drop was reduced to 11.8% compared to 20.2% in the untreated control.

Jafari *et al*., (2024)The fig tree yields differed from 4.956 Kg $tree^{-1}$ in cow manure to 8.125kg $tree^{-1}$ in turkey manure. On the other hand, total soluble solids (TSS) were significantly elevated by Bio Health fertilizer before the fruit was stored. Moreover, the partridge manure considerably increased the amount of gallic acid in the fruit (6.11 mg g DW).

## 2.3 Foliar Feeding

The foliar feeding method is one of the successful methods, but it cannot eliminate the importance of the roots in absorbing nutrients from the soil solution, (Abdul, 1988), indicated that foliar feeding can cover 85% of the plant's needs, so it is complementary to soil additions and not a substitute, Accordingly, foliar feeding by spraying the green parts of the plant with diluted solutions of nutrient salts several times is one of the important and successful methods for treating element deficiency, especially micronutrients and, to some extent, major nutrients.

Foliar application is one of the methods in agriculture, as this technology has spread widely in Iraq for various crops and fruit trees, and it was based on adding fertilizer by spraying it on the green group in the form of a liquid solution and in concentrations that are not harmful or deforming to the plant tissue, and it is no less efficient than absorbing nutrients through the roots (Tahir and Hassan 2005).

Foliar feeding is one of the important signs of modern agricultural development methods, as research and experiments have proven the possibility of supplying plants, fruit trees, and other crops with various nutritional elements by spraying plants with solutions of these elements, which the plant absorbs leaves, in addition to different plant parts that appear above the soil surface, such as stems and fruits (Crouch and Vanstaden, 2005; Taiz and Zeiger, 2006).

## 2.4 Physiological Effect of Calcium.

Many researchers have highlighted the importance of nutrients in the growth and production of crops, as the effects of these elements extend beyond just plant growth and development, and they influence various vital interactions in plant tissues, which can impact the yield and composition of proteins, fats, carbohydrates, and vitamins (Rashidi, 1987). Calcium is a critical element in the structure of cell wall tissues, enhancing their resistance to decomposition by enzymes (Galacturonase) that can cause fruits to soften during ripening. Additionally, it minimizes the breakdown of cell walls caused by microbial enzymes that infect the fruits, subsequently reducing their marketability and storage quality (Poovaiah *et al*., 1988). A deficiency of the necessary element hurts plant growth or vital processes, so it has become necessary to provide these elements by spraying them on the green group to be absorbed by the plant's green tissues directly to avoid the fixation and washing processes they are exposed to when added to the soil (Al-Sahaf, 1989).

Calcium is one of the major minerals (Macronutrients) that have many physiological functions in plant growth and development. Calcium enters into the structure of the plant, where it forms with pectic acid calcium pectate, which is a component of the middle plate of the plant cells, it is believed that calcium is important in the formation of cell membranes, and therefore calcium is necessary for both the cell wall and plasma membranes to perform their functions normally, Calcium is also important because it controls fruit cracking; although it is a relatively slow-moving element, its results have proven that the fruit content increases when spraying or immersing the fruits with calcium, and in the case of calcium deficiency below the minimum level, it leads to fruit cracking, and this is due to the divalent nature of calcium that increases the strength of the cell walls inside the fruits by forming pectate that binds with calcium and increases the resistance of the fruits to internal pressure (Siddiqui and Bangerth, 2004).

Using nanoparticles and nano powders can produce fertilizers with controlled or delayed release, Nanoparticles' high reactivity can be attributed to their larger specific surface area, higher density of reactive regions, or higher reactivity of these areas on the particle surfaces. These characteristics facilitate the easier absorption of nanoscale-generated fertilizers and insecticides. (Salama *et al*., 2014).

Nanotechnology opens up a wide range of new applications in the fields of biotechnology and agricultural industries because nanoparticles have unique physical and chemical properties, such as high surface area and high reactivity. Nanotechnology as "magic bullets" (Siddiqui *et al*., 2015).

In recent years, nanotechnology has spread to plant sciences and improved agricultural production, as nanomaterials are of physiological importance, as they are currently used to improve and increase the absorption of fertilizers by plant cells by reducing the loss of nutrients. The nano element has a high surface area, which enables it to efficiently deliver nutrients to the plant. It has also been shown that nano fertilizers improve the percentage of seed production, seed growth and germination, phosphatic activity, nitrogen and carbohydrate exchange, and protein synthesis in plants (Priyanka *et al*., 2015).

Calcium ions ($Ca^{+2}$), which are essential nutrients for plants, are crucial for the cell wall and membrane because they act as anions' compensation for both organic and inorganic anions in the vacuole, these ions control physiological processes such as the lengthening of root hairs, the formation of pollen tubes, and the movement of stomatal guard cells, and act as an intracellular messenger within the cell. (Pauly *et al*., 2000; Gao *et al*., 2019).

It is well-known that calcium deficiency or excessive calcium concentration causes disorders in humans and plants. Calcium ($Ca^{+2}$) is an essential ingredient for both plant and animal cells; it is one of the most essential second messengers, and it is essential for adjustable signaling molecules in all eukaryotic species. (Ranty *et al*., 2016; Gao *et al*., 2019).

The exogenous application of calcium stabilizes the plant cell wall by maintaining tissue firmness and reducing weight loss in fruits. (Shiri and Ghasemnezhad, 2019). Modern nanotechnologies are becoming widely recognized in agriculture to enhance crop yields with a healthy agroecosystem under environmental adversities; future agriculture may be built on the usage of nano-enabled fertilizers in many ways, by reducing nitrogen losses from leaching and avoiding chemical changes, these nanoscale fertilizers improve environmental quality and nutrient usage efficiency. (Verma *et al*., 2022).

**2.4.1 Effect of Calcium on vegetative growth traits:**

Lara *et al*., (2004) reported that calcium treatment prevents the loss of chlorophyll in plants and the dissolution of the middle plate of the outer cortex.

Hui *et al*., (2009) concluded that when studying the effect of spray calcium treatment on the vegetative growth characteristics of apricot fruits when using three concentrations of 0, 1, and 3% of calcium, the 3% concentration was superior in the studied characteristics.

Ali *et al*., (2019) reported that foliar application of calcium at 3 g $L^1$ on grapevines significantly increased leaf dry weight and chlorophyll content. Leaf dry weight increased from 1.84 g in the control to 2.63 g, and total chlorophyll content rose from 1.96 mg $g^1$ to 2.85 mg $g^1$. These effects were related to enhanced membrane stability and metabolic activity.

Mohamed *et al*., (2020) found that foliar spraying of nano-calcium at 3 g $L^1$ on fig trees significantly enhanced vegetative growth. The treatment increased leaf area from 68.4 cm² in the control group to 85.7 cm², and the total chlorophyll content rose from 2.11 mg $g^1$ fresh weight to 3.02 mg $g^1$. These improvements were attributed to calcium's role in stabilizing cell walls and enhancing photosynthesis.

Mustafa *et al*., (2022) observed that foliar application of nano-calcium-containing fertilizers at a concentration of 300 ppm significantly increased leaf area (from 68.6 to 85.4 cm²) and leaf dry weight in fig trees (*Ficus carica* cv. Black Mission). Similarly, chlorophyll content was markedly improved at 400–500 ppm. These effects were attributed to calcium's role in promoting cell division, membrane stability, and chlorophyll biosynthesis.

Otieno and Mwangi (2023) reported that foliar spraying of $CaCl_2$ at a 2% concentration significantly enhanced vegetative traits in fig trees. Leaf area increased from 62.5 to 78.9 cm², leaf dry weight from 4.35 to 5.87 g, and total chlorophyll content from 2.05 to 2.98 mg g$^1$ fresh weight. Additionally, leaf nitrogen concentration rose from 2.15% to 2.62%. These improvements were attributed to calcium's role in stabilizing cell membranes, activating photosynthesis, and enhancing nitrogen metabolism.

According to studies, Özdemir and Bayındırlı (2025). Two-season trial (2015–2016) on 'Sultani' figs applied foliar sprays of CaO (2.5, 5 mL L$^1$). The 5 ml L$^1$ Ca achieved the highest values in leaf dry weight, leaf area, and nutrient content (N, P, Ca, B), along with. Calcium alone also enhanced fruit volume and fresh weight. Overall, foliar Ca boosts vegetative growth—especially under saline stress—by strengthening cell walls and metabolism.

## 2.4.2 Effect of Calcium on the qualitative and quantitative characteristics:

Spraying Nijtaka pear trees with calcium chloride at a concentration of 0.3% at periods of (50, 80, 110) days after full flowering led to a shift in the proportions of sugars during storage to fructose and glucose, as the proportion of fructose in the fruits was twice the proportion of glucose (Cheolku *et al*., 2000).

Hayat *et al*., (2003) reported that decrease in the percentage of total sugars in apple fruits when treated with calcium chloride at a concentration of 2%, as the percentage of total sugars in the fruits treated with calcium was 10.31%, while the highest percentage of sugars was in the control treatment, reaching 11.77%, and the total soluble solids content of apples treated with 2% calcium chloride was 00.14%, while the total soluble solids content of the control treatment was 16.46%.

In another study, spraying pear trees with calcium chloride did not have a significant effect in increasing the percentage of total and reducing sugars at this stage of maturity, but this percentage increased insignificantly with the increase in calcium level, as the 2% calcium chloride treatment gave the highest rate of this percentage, while in the control treatment and for the same storage period. As for reducing sugars, they increased after the first three weeks of storage, and this increase continued after six weeks of storage. The 2% calcium chloride treatment was superior in giving the highest rate of reducing sugars percentage, respectively, while the percentage of reducing sugars in pear fruits decreased for the comparison treatment, respectively (Al-Mufarji, 2006).

Al-Anabi, (2008) also indicated through his study on the fruits of the Black Diyala fig variety that treatment with 2% calcium chloride led to a reduction in the percentage of total soluble solids compared to the control treatment, and when fig trees were sprayed with concentrations of calcium chloride and Vapor-Guard. Their effect was not significant on the percentage of total neutralizable acidity in fig fruits at maturity and the end of the storage period.

Showed study by Cesur and Ulusaran (2008) impact of spray calcium on fig trees. Recorded that fruit moisture with treated fruits containing about 75–78% moisture compared to 70–72% in controls, and reduced the rate of fruit drop from around 15% in untreated trees to 7–8% in treated ones. Calcium helps maintain fruit moisture content, preventing excessive water loss that leads to shriveling.

Khalifa *et al*., (2009) reported using calcium chloride at concentrations of 0.0, 0.2, and 0.4% through spraying in two periods. The first was after the fall of the plant corolla and the second after the fruit set - where the results showed that spraying with calcium chloride significantly increased the characteristics of the yield and the fruit yield in apple trees of the (Anna) variety and improved the physical and chemical characteristics of the fruits, as well as improving the nutritional status of the trees and reducing the severity of flower fall compared to trees not treated with foliar spraying.

In another study conducted by Ali *et al*., (2014) on pear trees on the effect of spraying calcium chloride on peach trees at two levels, namely 1 and 2%, it was found that the concentration of 1% was significantly superior in the studied traits, which are fruit weight and fruit diameter, and reduced the percentage of falling. As for the chemical traits of the fruits, there were no significant differences between the concentration of total sugars %, soluble solids, and total acidity.

The results showed in a study conducted by Duane, (2017) the effect of calcium prohexadione on the fruit set percentage and size of apples Malus domestica L. when using three concentrations of calcium prohexadione, namely 0, 125, and 250 mg $L^{-1}$, was found that the concentration of 125 mg $L^{-1}$ was superior in the studied traits represented by (number of fruits, fruit weight, increased number of flowers, and fruit size) compared to the control treatment.

The highest fruit weight loss was 3.9%, recorded when fig plants were sprayed with 0.5% Ca ($NO_3$), and fruits remained hard when plants were sprayed with 0.5% Ca $(NO_3)^2 + 30$ mg $L^{-1}$ NAA after storage for 15 days. Spraying plants with calcium enhances fruit hardness and prolonged fruit storage, because the calcium promotes cell wall formation (Zhang *et al*., 2019a).

In the study by Oliveira *et al.*, (2023), Foliar application of calcium compounds, such as calcium chloride ($CaCl_2$), significantly increases fruit weight and volume, for example, fruit weight increased to approximately 120–130 g compared to 90–100 g, in control, while fruit diameter reached 40–45 mm, in controls, 30–35 mm.

## 2.5 Storage Characteristics

Fruit storage is a vital process that has long attracted the attention of researchers due to its essential role in maintaining crop quality and reducing spoilage. Despite the progress made in studies related to this field, there are still fundamental challenges facing researchers and producers alike, intending to ensure the safe storage of fruits while maintaining their nutritional value for the longest possible period, whether in refrigerated or non-refrigerated storage conditions. Recent studies have addressed the effect of treating fruits with calcium chloride in enhancing the storability of fruits, in addition to determining the optimal temperatures and appropriate storage duration under these conditions, Despite the slow nature of calcium movement within plants, research results have confirmed that spraying plants with calcium chloride leads to an increase in the content of this element in leaves and fruits; it is noted that calcium deficiency below critical levels leads to the emergence of physiological damage this is one of the most important causes of fruit spoilage and a decrease in its quality during storage. This is attributed to the role of divalent calcium in enhancing the hardness of cell walls inside fruits through the formation of calcium pectin compounds, which increases the resistance of fruits to mechanical stresses and internal pressure during the storage period (Abbas, 1987). The water content of the fruits is the most important storage characteristic, as the loss of water from the fruits leads to a decrease in the cell filling pressure, and then the fruits wither, which has a significant effect on the marketing value of the fruits. The loss of water also has a significant effect on the weight of the fruits (Shirikvo, 1988). Refrigerated storage led to a reduction in fruit spoilage when compared to fruits stored at room temperature. The reason for the increased percentage of fruit spoilage may be due to physiological and pathological damage. The temperature and length of storage also affect the percentage of spoilage, as the percentage of spoilage increases with the rise in temperature and the increase in the length of the storage period (Lateef, 2022).

## 2.5.1 Effect of Organic Fertilizer on Storage Characteristics:

Marfaing and Poupot, (2017) The effect of using 0.5 % Ascophyllum nodosum extract this is one of the organic fertilizers (Seaweed Extracts), on pre-harvest fig trees improved post-harvest fig storage at 6 °C. Fruit from treated trees showed 28 % less weight loss after 10 days, 20 % firmer texture, and 30 % lower decay (Penicillium spp.) than untreated controls, attributed to enhanced cuticular wax and induced defense proteins.

Özdemir and Bayındırlı, (2018) Effects of adding humic acid (0, 500, 1,000, 1,500 mg L¹) were applied to 'Brown Turkey' figs pre-harvest. After harvest, the fruit was stored at 4 °C for 14 days. The 1,000 mg L¹ treatment reduced weight loss by 22 %, maintained firmness 18 % higher than control, and slowed total soluble solids increase, indicating delayed ripening. Decay incidence dropped from 35 % in controls to 12 % under humic acid, likely via enhanced phenolic defenses.

Wang et al., (2019). Discovers effect of fertilizer contain Bacillus subtilis bio film suspension adding on fig fruit before cold storage (4 °C) reduced Penicillium decay by 65 %, maintained firmness (+20 %), and lowered weight loss (–18 %) over 15 days—owing to bacterial antagonism and induced fruit defense enzymes (β-1,3-glucanase, chitinase).

Jawad et al., (2025) showed in a 2024 field study on Ficus carica 'Black Diyala' that spraying calcium oxide (5 mL L¹) significantly improved fruit quality characteristics both pre- and post-harvest. Treated fruits also recorded higher vitamin C content (7.26 mg/100 g juice vs. ~6.3 mg) and total soluble solids (18.96% vs. 16.34%). After cold storage (0–5 °C, 85% RH for 3 weeks), weight loss was significantly reduced in treated fruits (from 8.5% in control to 5.2%), and the percentage of non-marketable fruits (those showing decay or softening) was also reduced.

## 2.5.2 Effect of Nano Calcium on Storage Characteristics:

Spraying Nijtaka pear trees with calcium chloride at a concentration of 0.3% at periods of (50, 80, 110) days after full flowering led to a shift in the proportions of sugars during storage to fructose and glucose, as the proportion of fructose in the fruits was twice the proportion of glucose (Cheolku *et al*., 2000).

Al-Hamidawi (2002) concluded that by treating local apple fruits with three concentrations of calcium chloride (2, 4, 6) and then storing them for three months at a temperature of 5 °C, the percentage of weight loss decreased compared to the control treatment.

Hayat *et al*., (2003), in their study, discovered that the weight loss percentage of apples treated with 2% calcium chloride decreased, while the weight loss percentage of apples in the control treatment increased after a storage period of 60 days.

Another study showed that treating a portion of Shahmiveh pear fruits with 2% calcium chloride and the other portion with 2% calcium chloride plus 100 mg/L chlorine, after a storage period of 10 weeks at 2°C and 85% relative humidity, resulted in a weight loss of (30-40) % for the two treatments, respectively, compared to fruits not treated with calcium chloride, in which weight loss rates increased (Ashtari, 2004).

Fonseca (2005) also showed that spraying with calcium chloride before harvest affected the productivity and quality of fruits after harvest, as the weight of the fruits increased by 10% and their hardness increased at harvest, but the total soluble solids were lower in the fruits treated after 21 days of storage at a temperature of 5-7°C. Thus, the researcher explained that calcium increases crop production under environmental stress conditions.

Spraying with different concentrations of calcium chloride led to a reduction in the percentage of weight loss in pear fruits, as the concentration of 2% of calcium chloride had an effect in reducing the weight loss rates throughout the storage period after 3 or 6 weeks of storage, respectively, while the control treatment led to an increase in the weight loss percentage for the same period (Al-Mufarji, 2006).

Al-Mayahi and Abbas, (2006) showed that spraying with calcium significantly reduces the percentage of total soluble solids in jujube fruits.

Abd El-Motty *et al.*, (2007) in their study on the effect of pre-harvest calcium and boric acid treatment on the storage capacity of seven-year-old Canino apricot fruits stored for fifty days at a temperature of (0.5°C) showed a significant increase in weight loss and ascorbic acid loss, as well as an increase in total soluble solids, which led to a reduction in sugar losses and preservation compared to the control treatment.

Al-Mayahi (2007) stated that spraying calcium on Ziziphus trees of the olive and Bombawy cultivars led to a high decrease in the percentage of fruit cracking and an increase in the percentage when the added concentration was increased in both cultivars. Spraying with calcium solution led to a reduction in the damage of cracking of the fruits, as the percentage of cracking decreased in the calcium treatment at a concentration of 2000 mg $L^{-1}$ and in the treatment at a concentration of 1000 mg $L^{-1}$ for the olive and bomba varieties, respectively.

Treating fig fruits with calcium reduced the percentage of weight loss of the fruits when stored at a temperature of 5 °C compared to the control treatment, which had the highest percentage of weight loss of 5.10 and 82.4% (Al-Anabi, 2008).

Irget *et al.*, (2008) The maximum fruit weight of 17.14g gained at 280g Ca +NPK in Kanlıbahce and 70g Ca in Erbeyli was 16.48g. The highest yields of the fig tree were obtained with NPK application as 30.87 kg tree$^{-1}$ in Kanlıbahce and NPK + 280g Ca submissions in Erbeyli as 20.92 kg tree. The maximum marketable yields were in NPK treatment in Kanlıbahce, 26.91 kg tree$^{-1,}$ and NPK + 280g Ca 16.57 kg tree$^{-1}$ treatment in Erbeyli,

Shirzadeh *et al.*, (2011) in their study on immersing Jonagold apple fruits in different concentrations of $CaCl_2$ (0%, 2%, and 4%) and storing them for various periods (20, 40, 60, 80, 100, 120, and 150 days) reported a significant increase in fruit firmness and titratable acidity (TA) in the calcium-treated fruits compared to untreated ones. Additionally, calcium treatment significantly reduced weight loss, the TA/TSS ratio, and ethylene production in the fruits.

Abdrabboh (2012) concluded that giving a level of calcium 1.5% was beneficial in maintaining membrane integrity and reducing the leakage of phospholipids, proteins, and ions, which led to a decrease in the weight of apricot fruits.

Jan *et al*., (2013) found that immersing fruits from three apple cultivars (Mondial Gala, Royal Gala, and Golden Delicious) in two concentrations of $CaCl_2$ (0% and 9%) for 12 minutes and storing them for 150 days at 5 ±1 ºC and 60-70% relative humidity significantly increased fruit firmness and reduced weight loss compared to untreated fruits.

Hadi and Al-Shammari, (2013) observed that immersing Zaginya apricot fruits in a 5% calcium chloride solution for five minutes, followed by packaging in polyethylene bags and storing at 4 ±1 ºC and 80-85% relative humidity for one week, resulted in significant preservation of total soluble solids (TSS), titratable acidity (TA), and a reduction in weight loss compared to untreated fruits.

Sinha *et al*., (2019) found in their study that foliar treatment with 0.5% calcium chloride increased some chemical and biological properties, vitamin C, and fruit quality of the apricot variety 'Jahangiri'. Foliar spraying with calcium salt during fruit development is a safe method to replenish endogenous calcium in fresh fruits.

The fig trees sprayed with $CaCl_2$ caused a decrease in fruit weight loss. Fruit weight loss reduction is about 20.7% with the use of 0.9% $CaCl_2$. The fruit decay decreased with the increase of $CaCl_2$ concentration; 66.8% reduction obtained at 2.0% $CaCl_2$. Furthermore, the highest fruit firmness was 0.45N at the use of 1.0% $CaCl_2$ increased by about 59.8% compared to the control, as well as increased fruit diameter. The interaction between 0.1% $CaCl_2$ with 0 days of storage gave the highest fruit total soluble solids, which was about 13%. However, total acidity decreased at the initiation of storage compared to the control, and after that increased and reached about 0.15% at the use of 0.5% $CaCl_2$ and 21 days of storage. The total phenol in fig peel was about 440 and 450mg GAE 100 $g^{-1}$ obtained with 1.0 and 2.0% $CaCl_2$ at harvest (day 0) reduction throughout the storage, with the lowest value at the control. The highest antioxidant activity was about 100mg TEAC 100 $g^{-1}$ at the 1.0% $CaCl_2$ at harvest, reduced throughout the storage, and the lowest value was at the control.(Souza *et al*., 2023).

### 3. Materials and Methods

#### 3.1 Location of the experiment :

This experiment was conducted under plastic house conditions at one of the private farms in the village of Kani Sard in the Sharbazhir district near the Sitak area, which is about 35 km northeast of the center of Sulaymaniyah Governorate – Iraq (latitude: 35°38'25.7"N, longitude: 45°35'12.0"E, altitude). The experiment was conducted from 1/ April / 2024 to 31/ December / 2024. The plastic house measures 52m in length by 9m in width and 3m in height. It was covered with polyethylene film that had a thickness of 200 μm. Additionally, the study site was created using this plastic film.

#### 3.2 Soil Analysis :

Five soil samples were taken on 1/March/2024 before the start of the research in the shape of the letter X in five areas in the experimental field (plastic house) and at a depth of 0-30 cm to conduct chemical and physical analyses. Laboratory analyses of the field soil were conducted in the laboratories of the Directorate of Agriculture Research, Bakrajo in Sulaymaniyah Governorate, as shown in **Table (1).**

**Table (1):** Some physical and chemical properties of the soil were included in the research.

| Type of analysis | Rate | Unit |
|---|---|---|
| E C | 1.22 | ds.m$^{-1}$ 25 ℃ |
| pH | 8.04 | |
| Available N | 0.46 | % |
| Available P | 317.795 | Ppm |
| Soluble K$^+$ | 0.74 | meq/ L |
| Soluble Na$^+$ | 3.39 | meq /L |
| Soluble Ca$^{+2}$ | 8 | meq / L |
| Soluble Mg$^{+2}$ | 0.5 | meq/ L |
| CI | 0.5 | meq / L |
| HCO3$^-$ | 1.5 | meq/ L |
| CO$_3$$^{-2}$ | 0 | meq/ L |
| O.M | 3.06 | % |
| C | 1.78 | % |
| CaCo$_3$ | 20 | % |
| sand | 32.3 | % |
| Silt | 27.7 | % |
| Clay | 40 | % |
| Texture | Clay loam | |

### 3.3 Plant materials:

Using the cordon method, 54 uniform three-year-old fig trees were planted at a distance of 2×3 meters and used in this study view **(Figure 1).**

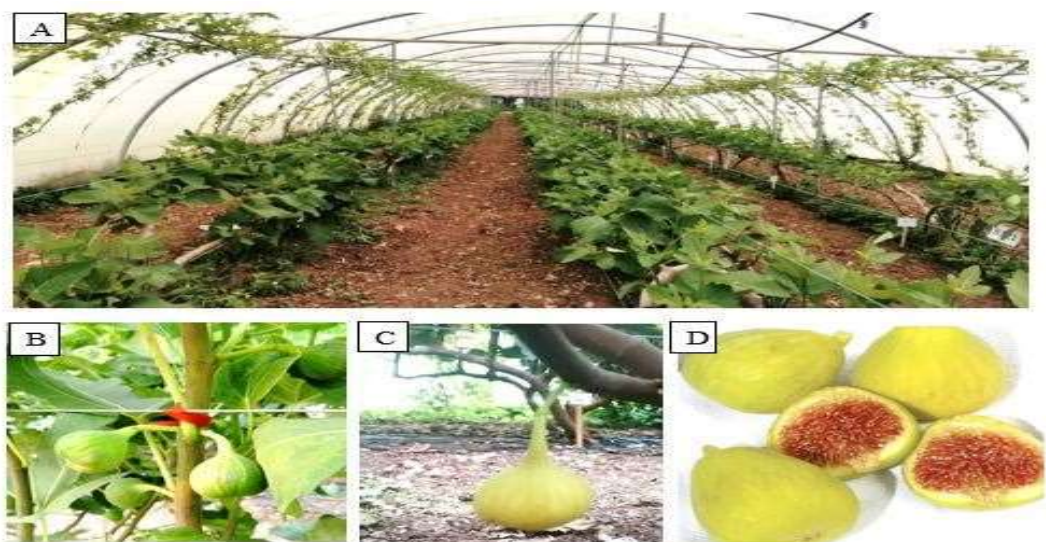

**Figure 1:** **A:** A fig tree using the cordon method under a plastic house
**B**: A branch fig tree with immature fig fruits. **C** and **D**: Mature fig fruits as a sample unit.





### 3.4   Treatments of Experiment

This study was designed with the following treatments to determine the effects of organic fertilizer (Bio Health) and calcium nanoparticle fertilizer alone and in combination on the quality and quantity development of fig fruit plants under plastic house conditions:

The first factor (B) is Bio Health organic fertilizer per tree at three levels: B1 is Control (0 g tree$^{-1}$); B2 is (15 g tree$^{-1}$); B3 is (30 g tree$^{-1}$).

The second factor (C) is Calcium nanoparticles at three concentrations, C1:  sprayed with distilled water; C2: (75 mg L$^{-1}$); C3: (150 mg L$^{-1}$).

Bio Health organic fertilizer was added into the soil using three levels per tree (0, 15, and 30 g tree$^{-1}$), in addition, Nano Calcium was sprayed at three levels per tree (0, 75, 150 mg L$^{-1}$) by a 16-liter backpack sprayer until completely wet. The fertilizer was added in two batches during the growing season with sprayed in three batches including; the first added (16/ April/ 2024) and the second (16/May/ 2024), And the first spraying after a week of fruit holding (25/May/ 2024), the second spray after 15 days of the first one, and the third after 15 days of the second spraying.

Table (2) Components of Bio Health fertilizer. It is an environmentally friendly product that enhances soil fertility by activating beneficial microorganisms within the soil.





Table 2: The contents of the Bio Health fertilizer

| Contents | Amount |
|---|---|
| Seaweed Extract | 10 % |
| Trichoderma harzianum | $10^6$ cfu/g |
| Bacillus Amyloliquefaciens | $10^7$ cfu/g |
| Potassium Humates | 70 - 75 % |
| Total Humic Acids [½] | 65 - 66 % |
| Humic Acid [1] | 47 - 49 % |
| Humic Acid [2] | 62 - 63 % |
| Fulvic Acid [1] | 2 - 3 % |
| Potassium ($K_2O$) | 10 - 12 % |
| Dry Matter | 83 - 85 % |
| Organic Substance | 65 - 69 % |
| Ph | 9.5 - 10.5 |
| Bulk Density | 0.55 - 0.65 g/cm$^3$ |

## 3.5 Agronomic practices and Treatments application

The common agricultural practices necessary inside the plastic house were performed, such as plowing, weed control, irrigation, thinning, pruning, and controlling diseases and insects as harmful as essential using natural anti-disease methods. Soil application and foliar spraying methods were performed for all treatments.





**3.6     Data Analysis:**

A randomized complete block design (R.C.B.D.) with two factors within the factorial experiments and three sectors was designed and laid out to set up the experiment. The treatments were distributed randomly within each sector at a rate of two trees for each experimental unit, so that 18 trees were selected as one replicate of plants, which was consequently 54 trees obtained for the total experimental unit. The variance was analyzed statistically and according to the unidirectional analysis table (ANOVA) using computer software according to the statistical analysis system program (SAS, 2001). The averages were compared using Duncan's Multiple Range at a probability level of 0.05 according to (Mead, R., and Hasted, 2003).

**3.7     The studied traits:**

**3.7.1 Vegetative characteristics:**

**3.7.1.1 Leaf area(cm$^2$):**

The measurements were taken at the end of August by taken Five fully grown and wide leaves from the main fruitful branches were taken from each experimental unit, starting from the third leaf to the sixth leaf from the top of the growth, using a computer program (J image) according to the method (AL-Obaidy *et al*., 2015).

**3.7.1.2   Total chlorophyll content (mg g$^{-1}$ fresh weight):**

The relative chlorophyll content in fully expanded leaves is estimated were taken at the end of August when harvested in different directions of the trees by taking 0.25 g of leaves and adding 15 ml of 96% ethanol for 24 hours in the shade, and then repeating the process twice in the same way for a total of 72 hours. Each package is read with 2 wavelengths (649 A and 665 A) using a





Spectrophotometer device, and then both chlorophyll A and chlorophyll B are calculated using the following mathematical equation (Knudson *et al*.,1977).

Chlorophyll a (mg g$^{-1}$ mw) = (13.70) (A665) – (5.76) (A649)

Chlorophyll b (mg g$^{-1}$ mw) = (25.80) (A649) – (7.60) (A665)

Total chlorophyll (mg g$^{-1}$ fresh weight) = chlorophyll a + chlorophyll b

A = wavelength (nm). [The chlorophyll concentration was given as µg Chlo mg dry weight.]

### 3.7.1.3    Leaf dry weight (g):

At the end of August, five leaves from each experimental unit were picked from the middle of the branches and rinsed thoroughly with plain water before being washed multiple times with distilled water to eliminate any dust or pesticides that had accumulated on them. Then, they were thoroughly dried with a cloth, and the wet weight was put in paper bags. An electric oven at a temperature of 70±5 °C was then used to dry samples till the weight was consistent. The leaves were then removed and weighed with a sensitive scale, and the dry matter percentage of leaves was calculated as described by (Al-Sahaf, 1989).

Leaf dry weight = dry sample weight (g) / fresh sample weight (g) X100

### 3.7.1.4    Nutrient content of leaves (N.P.K) %:

In the third week of August, five fully grown leaves were collected from each experimental unit from the fourth to the sixth leaf from the tip of the new growths, i.e., from the fully expanded, newly mature, and physiologically active leaves. They were washed with regular water and then distilled water to remove any dust and pesticide residues. After drying, they were placed in perforated paper bags, and put in an electric oven (vacuum oven) at a temperature of 70±5 ºC until the weight was stable, Then it was manually





smoothed with a hand grinder, 0.4 g of them were weighed and digested using sulfuric acid $H_2SO_4$ first by adding 3 ml of the acid and then adding concentrated perchloric acid $HClO_4$ at a ratio of 1:3 for each of them in order. After diluting the digestion solution, the nutritional elements were estimated using the following methods (Al-Sahaf, 1989) :

1-The percentage of total nitrogen (N%)using the Micro-Kjeldahl method. (A.O.A.C.,2005).

2-The percentage of phosphorus (P%)using ammonium molybdate was measured using a Spectrophotometer,(Estefan *et al*., 2013).

3-The percentage of potassium (K%)using a Flame Photometer,(Estefan *et al*., 2013).

### 3.7.2  Quantitative qualities

At the beginning of September, samples were taken to determine the characteristics of the quantity of fruits.

### 3.7.2.1  Fruit weight (g):

Ten fruits from each experimental unit were randomly weighed. They were weighed with a sensitive balance, and the mean fruit weight for each treatment was determined.

### 3.7.2.2  Fruit volume (cm³):

Ten random fig fruits were selected for each experimental unit to measure the size after two months following the last spray, using the volume of the displaced water. Then, the average fruit size extracted for each treatment was calculated as follows:

Fruit Volume = volume water after immersion− volume of water before immersion

### 3.7.2.3  Average yield of a tree (kg):





The total yield is calculated by multiplying the average fruit weight by the number of total fruits for each replicate according to the following :

Total yield = fruit weight rate x total fruit number

### 3.7.2.4  The percentage of moisture in the fruit (%)

A certain weight of fruit is dried using an electric oven at 100 degrees Celsius. Until the weight is stable, the percentage is extracted according to what was mentioned by (Al-Sahaf, 1989) according to the following law:

$$\% \text{ fruit moisture} = \frac{\text{sample weight before drying} - \text{sample weight after drying}}{\text{sample weight before drying}} \times 100$$

### 3.7.2.5  The percentage of dropped fruit %:

The number of fallen fruits from each experimental unit was randomly calculated from 10/June/ 2024, to 10/ September/ 2024, and then the total number of fruits, which consists of fruits that have reached maturity and are harvested, and the remaining fruits on the tree, including the fallen fruits, is extracted from the following equation:

$$\% \text{ dropped of fruit} = \frac{\text{number of dropped fruits for treatment}}{\text{Total number of fruits}} \times 100$$

### 3.7.2.6  Fruit growth curve depending on diameter (mm):

The average diameter of 8 fruits identified on the tree is calculated for each experimental unit and in each replicate, 15/June/2024. Weekly readings are taken from the same identified fruits using Digital Vernier Calipers (Model: DMV-SL05, WORKZONE, Germany), and then the average diameter of the fruit is extracted for each treatment in (mm).

### 3.7.2.7  Fruit hardness before and after storage (Newton):

The degree of fruit hardness was measured by taking (8) fruits from each experimental unit using a standard hardness device (TEXTURE ANALYZER)





and then extracted the fruit hardness rate estimated at (N), using a probe with a diameter [(TA4/1000 Cylinder) (38.1mm D.20mm L]. Newton (N) was used to express values. Then, I extract the average fruit hardness estimated at (kg cm$^2$), and after converting (kg cm$^2$ to Newton) by applying this modifier,

$$\frac{\text{Device Reading}}{100} \times 9.8$$

device Standard:  TRIGGER:  100.0 g

DEFORMATION: 5.0 mm

SPEED:  10.0 mm/S

According to the method ( King *et al*., 2012).

### 3.7.3   Qualitative qualities and characteristics:

To prepare samples, ten fig fruits were randomly taken from each experimental unit, each consisting of two trees. After collection, the fruits were kept in a refrigerated box at 4 °C and transported to the higher education laboratory at the College of Agricultural Engineering Sciences at the University of Sulaymaniyah. The samples were then ground with liquid nitrogen and kept in the freezer.

### 3.7.3.1 Soluble sugar content before and after storage (SSC) :

Soluble sugar content was estimated following the method defined by (Tahir *et al*., 2022 ). 0.1 g of ground fig fresh fruit tissue was added to 1000 µL of distilled water and shaken for 20 min. The samples were boiled at 92 °C for 30 min, cooled with cold water, centrifuged for 12 min at 8000 rpm, and the supernatant was collected. Anthrone reagent was prepared by dissolving 0.41 g of anthrone in 44 mL of distilled water and then adding 231 mL of $H_2SO_4$. 10 µL of the supernatant was mixed with 2000 µL of anthrone reagent. The solution mixture was incubated at 95 °C for 4 min, and the color of the solution changed to dark green. The solution of samples was cooled and read against the blank (anthrone reagent solution) at 620 nm, and a UV-visible spectrophotometer





(UVM6100, MAANLAB AB, Sweden) was used. Soluble sugar content was calculated by the following formula:

$$SSC(\mu g \, g \, FW) = \frac{V}{W} \times C$$

Where V is the volume of extract (mL), W is the fresh weight of the fig sample (g), and C is the concentration of glucose obtained from the standard curve (Figure 2). A stock solution of standard compound (glucose) was prepared by adding 10 mL of deionized water to 10 mg of glucose to get a final concentration of 1 mg mL. A series of dilutions of glucose (0, 4, 10, 20, 30, 50, 80 µg) was prepared. Linear regression was observed between the absorbance values at 620 nm and the glucose concentrations.

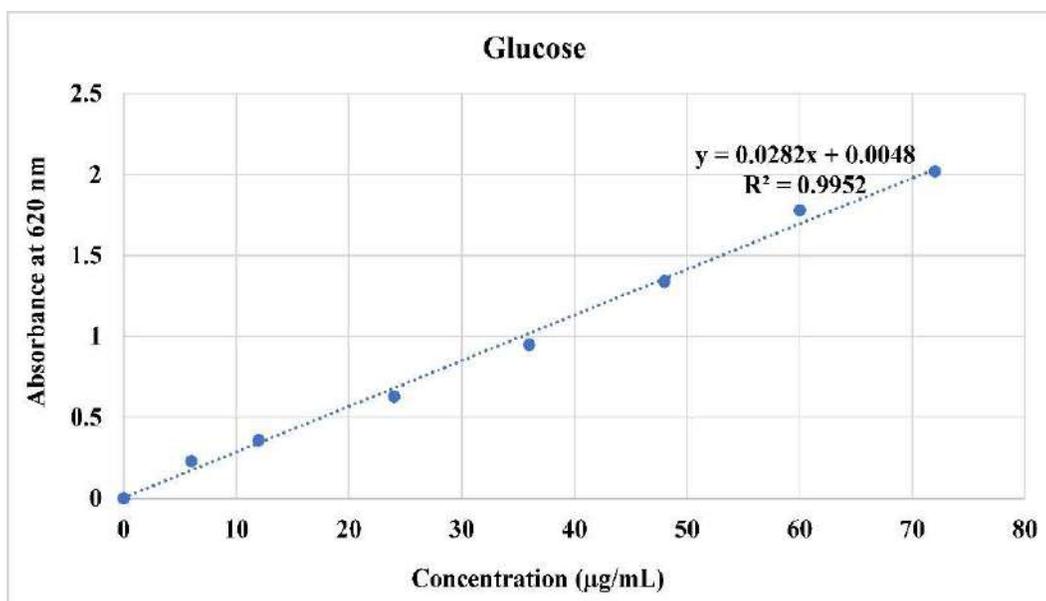

**Figure 2: Standard curve of glucose.**

### 3.7.3.2 Total phenolic content before and after storage (TPC):

According to the method (Maghsoudlu *et al.*, 2017), total phenolic content (TPC) was evaluated in fig fruit tissue. 0.1 g of ground tissue was mixed with





1000 µL of 80% (v/v) methanol and shaken for 40 minutes, then all samples were incubated overnight at 5 °C. The sample mixture was centrifuged at 12000 rpm for 15 minutes, and the supernatant was collected for TPC analysis. 150 µL of the supernatant was mixed with 1050 µL of 1:9 Folin–Ciocalteu reagent: water (v/v), after 7 min added 850 µL of 10% $Na_2CO_3$ was added and incubated in the dark for 30 minutes. After the reaction, the color of the mixture solution was changed to light blue and read at 750 nm against the blank (150 µL $dH_2O$ mixed with 1050 µL 1: 9 Folin–Ciocalteu reagent: water (v/v) and 850 µL 10% $Na_2CO_3$), a UV-visible spectrophotometer (UVM6100, MAANLAB AB, Sweden) was used. Gallic acid (GAE) was employed as a standard. The standard solution was prepared by dissolving 9 mg of gallic acid in 9 mL of methanol to attain a final concentration of 1 mg/mL. A sequence of dilutions of gallic acid (0, 50, 100, 150, 200, 250, 300 µg/mL) was used to produce a standard curve, and a linear association between the absorbance values at 750 nm and the gallic acid content was observed. The total phenolic content in each sample was determined using the standard curve (Figure 3). The following equation was used to calculate the TPC:

$$\text{TPC (µg GAE.g FW)} = \frac{v}{w} \times C$$

Where V is the volume of extract (mL), W is the fresh weight of the sample (g), and C is the concentration of gallic acid collected from the standard curve.





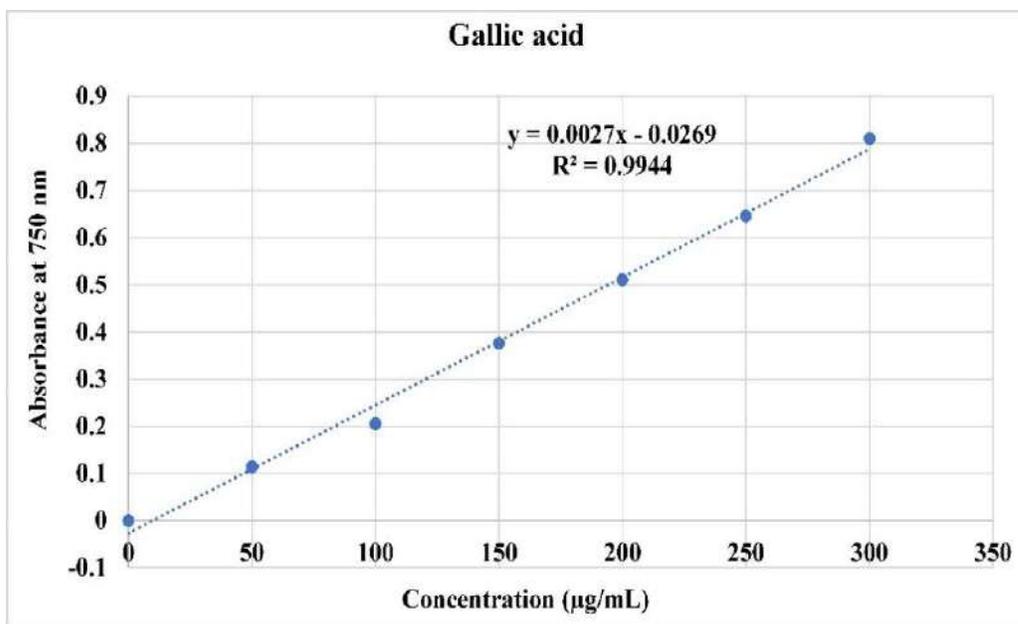

**Figure 3: Standard curve of gallic acid.**

### 3.7.3.3 Antioxidant capacity before and after storage (AC):

Antioxidant capacity was determined by using 0.1 g of ground fruit tissue, added into 1 mL of 80% methanol, and shaken for 40 minutes. Then all samples were incubated overnight at 5 °C, and the mixture was centrifuged at 12000 rpm for 15 min, and the supernatant was taken. 300 µL of the extract was mixed with 1.9 mL of 1-diphenyl-2-picrylhydrazyl (DPPH) solution (0.01g DPPH dissolved in 260 mL of 95% methanol). The sample mixtures were incubated in the dark for 30 minutes at room  temperature, and the samples were absorbed at 517 nm against the blank (95% methanol) using a UV-visible spectrophotometer (UVM6100, MAANLAB AB, Sweden) used (Tahir *et al*., 2024)

The standard compound, 6-hydroxy-2,5,7,8-tetramethylchroman-2-carboxylic acid (Trolox), was used to build the calibration curve. Trolox (12 mg) was combined with 12 mL of 80% ethanol (v/v) solvent and diluted to achieve concentrations of (0.00, 0.33, 0.66, 1.320, 2.00, 2.7, and 3.4 µg mL)





(Figure 4). Linear regression was found between the absorbance values at 517 nm and the varied Trolox concentrations. The following equation was used to estimate the antioxidant capacity:

$$\text{Antioxidant capacity by DPPH (µg Trolox g FW)} = \frac{V}{W} \times C$$

Where V is the volume of extract (mL), W is the fresh weight of the sample (g), and C is the concentration of Trolox determined from the standard curve.

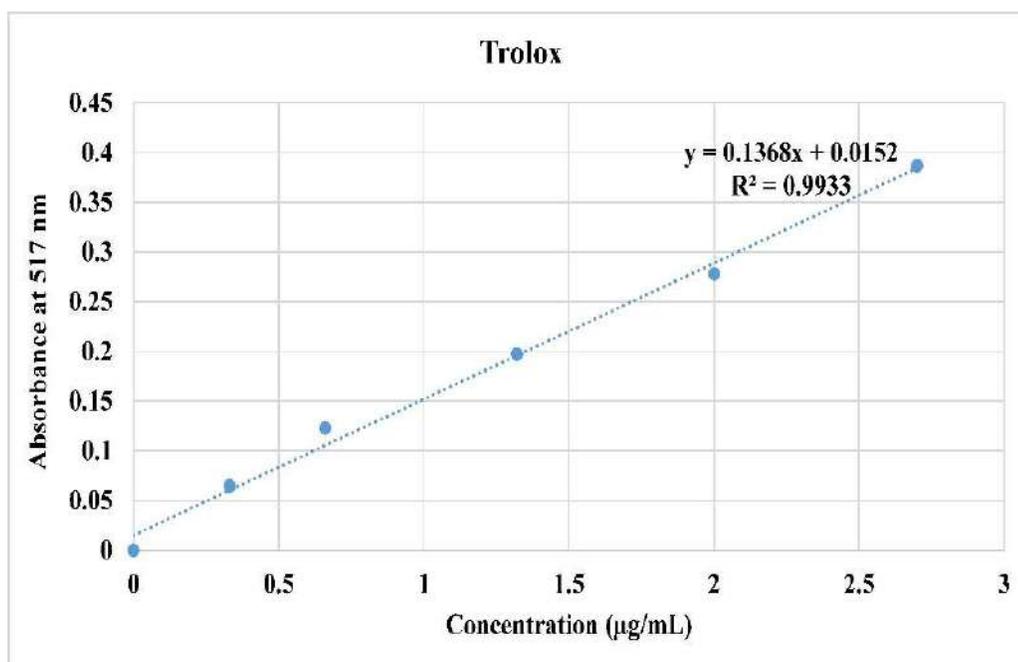

**Figure 4: Trolox standard calibration curve.**

### 3.7.3.4 Total flavonoid content before and after storage (TFC):

To determine the total flavonoid content (TFC), 0.1 g of dried fruit powder was extracted using 1000 µL of 80% (v/v) methanol. The mixture was incubated at 10 °C for 16 hours and then centrifuged at 14,000 rpm at 4 °C for 19 minutes. The resulting supernatant was collected for TFC estimation, following the method of Halshoy et al. (2024). A stock solution of quercetin (1 mg mL) was prepared by dissolving 11 mg of quercetin in 11 mL of deionized water. A standard curve for quercetin was established, with a linear regression between absorbance values at 415 nm and quercetin concentrations





(Figure 5). Concisely, 700 µL of the extract was mixed with 900 µL of 80% methanol, 300 µL of 2% (w/v) aluminum chloride, 80 µL of 1 M potassium acetate, and 1.7 mL of deionized water. The mixture was incubated at 28 °C for 30 minutes, and absorbance was measured at 415 nm using a UV-visible spectrophotometer (SHIMADZU, Japan). Each assay was conducted in triplicate, and mean TFC values were calculated to ensure accuracy.

$$\text{TFC (µg QE g FW)} = V/W \times C$$

Where V is the volume of extract (mL), W is the dry weight of the whole fruit (g), and C is the concentration of quercetin was determined from the standard.

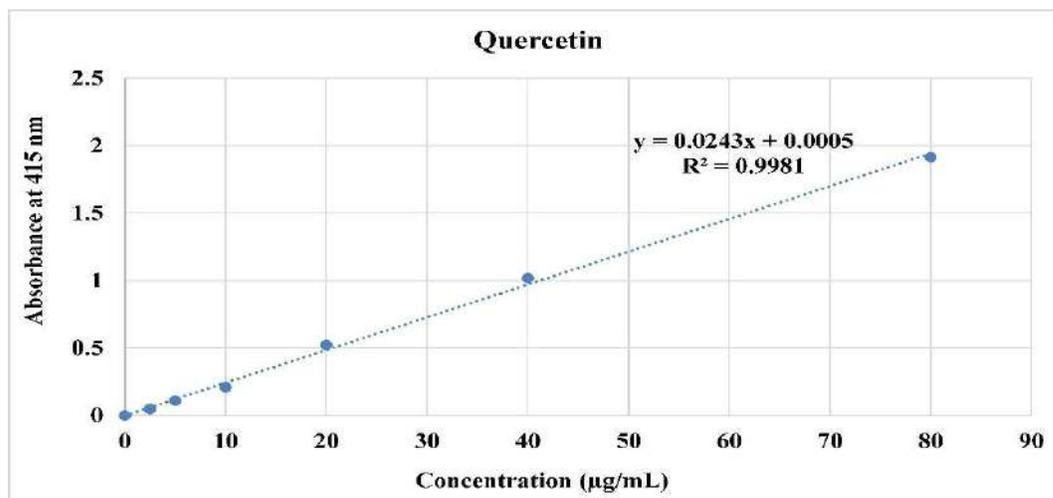

**Figure 5:** Standard calibration curve of quercetin

### 3.7.3.5 Ascorbic acid content before and after storage (ASA):

Ascorbic acid content (ASA) was determined by combining 0.5 g of powdered fig fruit tissue with 1300 µL of 1% (w/v) HCl and vigorously shaking the mixture for 30 minutes. The mixture was centrifuged for 10 minutes at 13000 rpm, and the supernatant was collected. The supernatant was mixed with 1000 µL of 1% (v/v) HCl and measured at 243 nm against a blank containing 1% (v/v) HCl. (Rasul, 2023).





The ASA was defined as µg/g of fresh flesh weight using the following formula:

$$\text{ASA}(\mu g.g \text{ FW}) = \frac{\text{Volume of juice (mL)}}{\text{Fresh weight of flesh (g)}} \times \text{Concentration standard curve ASA}$$

The standard curve of ascorbic acid was prepared by using 0, 5, 10, 15, 20, 25, and 30 mg mL of ascorbic acid (Figure 6).

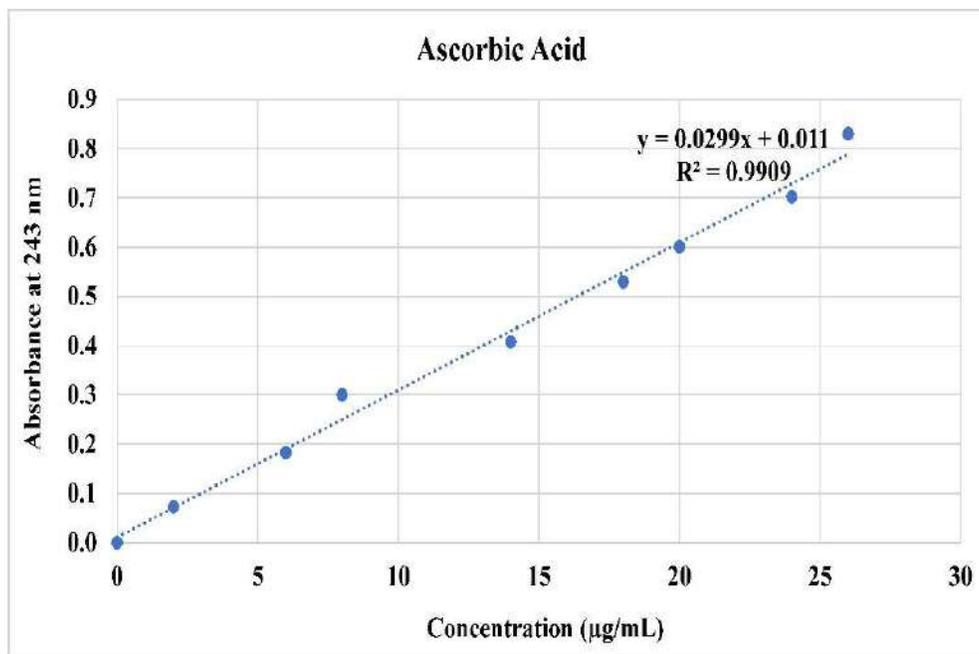

**Figure 6: Standard curve of ascorbic acid.**

### 3.7.3.6 Carotenoid content before and after storage (Carotd) :

One gram of powdered fig fruit tissue was mixed with 1000 µL of 100% methanol, and the mixture was incubated overnight at 5 °C. After centrifuging the samples for 8 minutes at 13000 rpm, 500 µL of the supernatant was collected and mixed with 1500 µL of 100% methanol. At 470 nm, the sample was read against a blank of 100% methanol. (Faraj,2023). The carotenoid concentrations were expressed as µg per gram of fresh flesh weight and estimated by this formula:

$$\text{Carotd} (\mu g \text{ g}^{-1} \text{ FW}) = \frac{\text{Absorbance reading x Total volume of juice (mL) x 10000}}{\text{Carotene extinction coefficient in methanol (2210) x Fresh weight of flesh (g)}}$$





### 3.7.3.7 Total carbohydrate content before and after storage (TCC):

The total soluble carbohydrate content was determined using the phenol-sulfuric acid method as described by ( Petkova *et al*., 2019). For the analysis, 0.1 g of fig extract was combined with 1 mL of 5% phenol and 5 mL of concentrated sulfuric acid. The mixture was then incubated in a water bath at 30°C for 20 minutes. After this incubation period, the absorbance was measured at a wavelength of 490 nm, with dH$_2$O used as the blank for calibration. The concentration of carbohydrates present in the fig extracts was quantified against a glucose calibration curve, and the results were expressed as μg glucose g fruit weight.

**Preparing the composite sample:** 10 ripe fruits were randomly selected for each experimental unit, which were harvested 3 months after the last spraying. The selected fruits were placed in an electric blender for 2-3 minutes, and the juice was filtered through a piece of cotton fabric. Then, all its qualitative and chemical yield characteristics were determined.

### 3.7.3.8 The percentage of total soluble solids in fruit juice before and after storage (% TSS):

A portable digital refractometer was used to measure the Total Soluble Solids (TSS) in fruit juice. Firstly, distilled water was used to adjust the refractometer. Then, three drops of the fruit juice were placed on the device's sensor, and the TSS was read.

### 3.7.3.9 The total acidity of fruit juice before and after storage (TA):

It was calculated on the basis that the dominant acid in the juice is citric acid. 0.5 ml of the juice sample was put into a small conical flask, mixed with drops of phenol naphthalene indicator, and then the titrated sample was





titrated with (0.1N) sodium hydroxide. TA was measured according to the following law:

TA %= (citric acid equivalent weight × titer × base volume)/ (1000× juice volume) ×100

### 3.7.3.10   The Ratio of Total soluble solids in fruit juice /the total acidity of fruit juice before storage

It was calculated by dividing the Total soluble solids (TSS) values by the Total acidity TA (%) values of the fruits (Taha, 2008).

### 3.7.4   Storage qualities:

The fruits of the first experiment with all of their treatments were picked at the same maturity stages and stored in the cold room at (0-5°C) and 85-90 ± % R.H. for 3 weeks After that, the fruits were taken from the cold room, at cold storage in the College of Agriculture / University of Kirkuk. The following parameters were taken from the fruits:

### 3.7.4.1. Percentage of Fruit weight loss (%):

After storing the fruits, each transaction has specific and known weights. The scheduled storage period, which is three weeks, passes, after which the fruit weights are calculated for each transaction after storage, and the percentage is calculated.

$$\% \text{ weight loss} = \frac{\text{Fruits weight at the beginning of storage} - \text{Fruits weight at the end of storage}}{\text{Fruits weight at the beginning of storage}} \times 100$$

### 3.7.4.2 Percentage of Fruit non-marketable (%):

The weight of the damaged fruits as a result of pathological and physiological injuries in the stored fruits was calculated as in the following equation:

$$\text{Percentage of total non-marketable fruits} = \frac{\text{weight of fruits is not marketable}}{\text{Fruit weight at the end of storage}} \times 100$$

## 4. 1 Result

### 4.1.1 Vegetative characteristics:

#### 4.1.1.1 Leaf area (cm$^2$):

Table (3) highlights the individual and incorporates the effects of Bio Health and Nano Calcium on the fig leaf area (cm$^2$). The use of 30g tree$^{-1}$ of Bio Health fertilizer recorded the highest value, 538.82 cm$^2$, and had significantly superior on the control was 421.02 cm$^2$, and 15g tree$^{-1}$ was 493.12 cm$^2$ . as well as 150mg L$^{-1}$ of Nano Calcium gave the highest value 507.73 cm$^2$ had a significant effect on the 75mg L$^{-1}$ was 481.40 cm$^2$, and control was 463.84 cm$^2$. The bilateral interaction of 30g tree$^{-1}$ Bio Health fertilizer and 150mg L$^{-1}$ of Nano Calcium recorded the highest value of 566.15 cm$^2$, superior to all the treatments. Still, the lowest value of leaf area was 402.53 cm$^2$ at the control.

Table 3: Influences of Organic Fertilizer, Nano Calcium, and their interactions on fig leaf area (cm$^2$) of figs.

The values with similar letters for each factor or their interactions individually did not differ significantly

| Bio Health g tree$^{-1}$ | Nano Calcium mg L$^{-1}$ | | | Average of Bio Health |
|---|---|---|---|---|
| | 0 | 75 | 150 | |
| 0 | 402.53 f | 411.18 f | 449.34 e | 421.02 c |
| 15 | 473.53 d | 498.14 c | 507.69 c | 493.12 b |
| 30 | 515.45 bc | 534.87 b | 566.15 a | 538.82 a |
| Average Of Nano Calcium | 463.84 c | 481.40 b | 507.73 a | |

according to Duncan's test at a  probability level of 0.05.

#### 4.1.1.2  Total chlorophyll content (mg g$^{-1}$ fresh weight):

The significant impacts of Bio Health fertilizer and Nano Calcium on the leaf's total chlorophyll content (mg g$^{-1}$ fresh weight) are shown in **Table (4)** as both single and interaction recorded significant effects. The highest value of leaf chlorophyll content was 22.52 mg g$^{-1}$ fresh leaf weight at 30g tree$^{-1}$ of Bio Health; however, the lowest value was 17.35 mg g$^{-1}$ fresh leaf weight at the control. Furthermore, using 150mg L$^{-1}$ of Nano Calcium obtained the highest total chlorophyll content, 21.13 mg g$^{-1}$ fresh leaf weight, but the lowest was 18.29 mg g$^{-1}$ fresh leaf weight at the control. The highest amount of chlorophyll in the leaf was 24.68 mg g$^{-1}$ fresh leaf weight at the interaction of 30g tree$^{-1}$ of Bio Health and 150mg L$^{-1}$ of Nano Calcium. While the lowest amount at the control was 15.46 mg g$^{-1}$ fresh leaf weight.

**Table 4.** Influences of Organic Fertilizer, Nano Calcium, and their interactions on total chlorophyll content (mg g$^{-1}$ fresh weight) of leaf figs.

| Bio Health g tree$^{-1}$ | Nano Calcium mg L$^{-1}$ | | | Average of Bio Health |
|---|---|---|---|---|
| | 0 | 75 | 150 | |
| 0 | 15.46 g | 18.02 f | 18.57 ef | 17.35 c |
| 15 | 18.80 ef | 19.56 de | 20.14 cd | 19.50 b |
| 30 | 20.61 c | 22.27 b | 24.68 a | 22.52 a |
| Average Of Nano Calcium | 18.29 c | 19.95 b | 21.13 a | |

The values with similar letters for each factor or their interactions individually did not differ significantly according to Duncan's test at a probability level of 0.05.

## 4.1.1.3 Leaf dry weight (g):

The statistical results shown in **Table (5)** illustrate the significant impact of Bio Health fertilizer and Nano Calcium as individual and in combination on the leaf dry weight. The highest value was 33.05g, achieved after utilizing 30g tree$^{-1}$ of Bio Health. However, the lowest value was 25.81g at the control. In addition, using 150mg L$^{-1}$ of Nano Calcium showed the highest value was 31.25g, and the lowest value was 27.26g at the control. The combined effect of 30g tree-1 of Bio Health and 150mg L-1 of Nano Calcium yielded the highest value at 37.59g, whereas the control recorded the lowest value at 24.03g of single leaf dry weight.

**Table 5.** Influences of Organic Fertilizer, Nano Calcium, and their interactions on leaf dry weight (g) of figs.

| Bio Health g tree$^{-1}$ | Nano Calcium mg L$^{-1}$ | | | Average of Bio Health |
|---|---|---|---|---|
| | 0 | 75 | 150 | |
| 0 | 24.03 f | 26.34 e | 27.06 de | 25.81 c |
| 15 | 27.67 de | 28.60 cd | 29.09 cd | 28.45 b |
| 30 | 30.09 bc | 31.48 b | 37.59 a | 33.05 a |
| Average of Nana Calcium | 27.26 c | 28.81 b | 31.25 a | |

The values with similar letters for each factor or their interactions individually did not differ significantly according to Duncan's test at a probability level of 0.05.

### 4.1.1.4 Leaf nitrogen content (%):

Meaningful effects of Bio Health and Nano Calcium on the leaf nitrogen percentage (%N) at both patterns, individual and interaction **(Table 6).** Bio Health fertilizer at a level of 30g tree-1 was recorded as having the highest value of leaf nitrogen content at 1.879%, which had a significant effect. This was superior to the 15g tree-1 level, which yielded 1.552%, and the lowest was 1.393% at the control.   In addition, using 150mg $L^{-1}$ Nano Calcium significantly proved that the highest nitrogen value was 1.678%. However, the lowest value was 1.546%. The binary influences of 30g $tree^{-1}$ Bio Health fertilizer and 150mg $L^{-1}$ Nano Calcium gave the highest value of leaf nitrogen content, which was 1.947%, while the lowest was 1.350% at the control.

**Table 6.** Influences of Organic Fertilizer, Nano Calcium, and their interactions on the leaf nitrogen percentage of figs.

| Bio Health g $tree^{-1}$ | Nano Calcium mg $L^{-1}$ | | | Average of Bio Health |
|---|---|---|---|---|
| | 0 | 75 | 150 | |
| 0 | 1.350 h | 1.390 g | 1.440 f | 1.393 c |
| 15 | 1.457 f | 1.553 e | 1.647 d | 1.552 b |
| 30 | 1.830 c | 1.860 b | 1.947 a | 1.879 a |
| Average of Nana Calcium | 1.546 c | 1.601 b | 1.678 a | |

The values with similar letters for each factor or their interactions individually did not differ significantly according to Duncan's test at a  probability level of 0.05.

### 4.1.1.5 Leaf phosphorus content (%):

The results of **Table (7**) clarified the considerable effects of Bio Health fertilizer and Nano calcium on the leaf phosphorus content (%P) in both individual and interaction shapes. The highest amount of phosphorus was 0.393% in the leaf at 30g tree$^{-1}$ of Bio Health.  On the other hand, using 150mg L$^{-1}$ of Nano Calcium gave the highest value of phosphorus, 0.366%. The interaction between both fertilizers also showed that the highest amount of phosphorus was 0.399% when using 30g tree$^{-1}$ of Bio Health and 150mg L$^{-1}$ of Nano Calcium. In contrast, the lowest value was 0.290% at the control.

**Table 7.**  Influences of Organic Fertilizer, Nano Calcium, and their interactions on the Leaf Phosphorus Percentage of figs

| Bio Health g tree$^{-1}$ | Nano Calcium mg L$^{-1}$ | | | Average of Bio Health |
| --- | --- | --- | --- | --- |
| | 0 | 75 | 150 | |
| 0 | 0.290 g | 0.317 f | 0.324 f | 0.311 c |
| 15 | 0.340 e | 0.357 d | 0.375 c | 0.357 b |
| 30 | 0.388 b | 0.393 ab | 0.399 a | 0.393 a |
| Average of Nana Calcium | 0.339 c | 0.356 b | 0.366 a | |

The values with similar letters for each factor or their interactions individually did not differ significantly according to Duncan's test at a  probability level of 0.05.

## 4.1.1.6 Leaf potassium content (%):

Leaf potassium content (%K) was significantly affected by Bio Health fertilizer and Nano Calcium individually and as a binary shown in **Table (8).** Using 30g tree[-1] of Bio Health fertilizer, the highest value of potassium was 1.052% compared to the control, which was 0.848% of leaf potassium content. On the other hand, the highest amount of potassium was 1.004%, achieved by 150mg L[-1] of Nano Calcium, but the lowest value was 0.897% at the control. Concerning the bilateral interaction between Bio Health fertilizer and Nano Calcium. The results also confirmed that the interaction between 30g tree[-1] of Bio Health and 150mg L[-1] of Nano Calcium recorded the highest amount of leaf potassium at 1.150%, outperforming all other interactions, with the lowest amount being 0.780% at the control.

**Table 8.** Influences of Organic Fertilizer, Nano Calcium, and their interactions on the leaf potassium (K%) of figs.

| Bio Health g tree[-1] | Nano Calcium mg L[-1] | | | Average Of Bio Health |
|---|---|---|---|---|
| | 0 | 75 | 150 | |
| 0 | 0.780 h | 0.863 g | 0.900 f | 0.848 c |
| 15 | 0.930 e | 0.953 d | 0.963 cd | 0.949 b |
| 30 | 0.980 c | 1.027 b | 1.150 a | 1.052 a |
| Average of Nana Calcium | 0.897 c | 0.948 b | 1.004 a | |

The values with similar letters for each factor or their interactions individually did not differ significantly according to Duncan's test at a probability level of 0.05.

## 4.1.2 Quantitative characteristics of fruit:

### 4.1.2.1 Fruit weight (g):

Considerable effects of the Bio Health fertilizer and Nano Calcium on fig fruit weight are shown in **Table (9).** The highest value of fruit weight was 104.76g at the use of 30g tree$^{-1}$ of Bio Health fertilizer, but the lowest weight was 75.51g at the control. In addition, using 150mg L$^{-1}$ Nano Calcium gave the maximum value of fruit weight of 96.55g as compared to the control, which gave the lowest value of 84.30g. The combination of both treatments revealed a significant effect, and the highest value was 110.13g at the 30g tree$^{-1}$ of Bio Health and 150mg L$^{-1}$ of Nano Calcium. At the same time, the lowest weight, 66.97g, was noted at the control**.**

**Table 9:** Influences of Organic Fertilizer, Nano Calcium, and their interactions on the fruit weight (g) of figs.

| Bio Health g tree$^{-1}$ | Nano Calcium mg L$^{-1}$ | | | Average of Bio Health |
|---|---|---|---|---|
| | 0 | 75 | 150 | |
| 0 | 66.97 g | 77.62 f | 81.93 e | 75.51 c |
| 15 | 85.85 de | 89.91 d | 97.58 c | 91.11 b |
| 30 | 100.07 bc | 104.07 b | 110.13 a | 104.76 a |
| Average of Nano Calcium | 84.30 c | 90.53 b | 96.55 a | |

The values with similar letters for each factor or their interactions individually did not differ significantly according to Duncan's test at a probability level of 0.05.

### 4.1.2.2 Fruit volume (cm$^3$):

Table (10) confirmed the significant effect of the use of Bio Health fertilizer and Nano Calcium on the fruit volume. The results documented that the maximum fruit volume was 115.03 cm³ after adding 30g tree$^{-1}$ Bio Health fertilizer, whereas the lowest fruit volume, 88.57 cm³, was achieved in the control. In addition, the use of 150mg L$^{-1}$ Nano Calcium indicated the highest fruit volume was 107.46 cm³, while the control yielded the lowest volume of 95.05 cm³. Regarding the combination of Bio Health fertilizer 30g tree$^{-1}$ and Nano Calcium 150mg L$^{-1}$, the largest fruit volume of 120.00 cm³ was verified. In comparison, the lowest fruit volume, 78.20 cm³, was obtained from the control.

Table 10: Influences of Organic Fertilizer, Nano Calcium, and their interactions on the fruit volume (cm$^3$) of figs.

| Bio Health g tree$^{-1}$ | Nano Calcium mg L$^{-1}$ | | | Average of Bio Health |
|---|---|---|---|---|
| | 0 | 75 | 150 | |
| 0 | 78.20 f | 91.33 e | 96.17 d | 88.57 c |
| 15 | 97.10 d | 98.67 d | 106.20 c | 100.65 b |
| 30 | 109.87 c | 115.23 b | 120.00 a | 115.03 a |
| Average of Nano Calcium | 95.05 c | 101.74 b | 107.46 a | |

The values with similar letters for each factor or their interactions individually did not differ significantly according to Duncan's test at a probability level of 0.05.

## 4.1.2.3 Average yield of a tree (kg):

The statistical analysis in **Table (11)** clarified that the application of Bio Health fertilizer and Nano Calcium for fig trees had a substantial influence on the average yield of a single tree. The 30g tree$^{-1}$ of Bio Health produced the highest yield, 10.436 kg tree$^{-1}$, significantly superior to the other concentrations, while the comparison of the control was recorded as the lowest yield, 7.687 kg tree$^{-1}$. Furthermore, the 150mg L$^{-1}$ of Nano Calcium gave the highest tree yield, 8.881 kg tree$^{-1}$, compared to the control, which recorded the lowest value of yield, 8.281 kg tree$^{-1}$. The interaction between the two fertilizers showed considerable effects. The highest value of yield was 11.113 kg tree$^{-1}$ obtained between 30g tree$^{-1}$ of Bio Health and 150mg L$^{-1}$ of Nano Calcium. The lowest weight average, 6.817 kg tree$^{-1}$, was noted in the control.

**Table 11**. Influences of Organic Fertilizer, Nano Calcium, and their interactions on The average yield of a tree (Kg) of figs.

| Bio Health g tree$^{-1}$ | Nano Calcium mg L$^{-1}$ | | | Average of Bio Health |
|---|---|---|---|---|
| | **0** | **75** | **150** | |
| **0** | 6.817 e | 7.783 d | 8.460 c | 7.687 b |
| **15** | 8.040 cd | 8.217 cd | 7.070 e | 7.776 b |
| **30** | 9.987 b | 10.207 b | 11.113 a | 10.436 a |
| **Average of Nano Calcium** | 8.281 b | 8.736 a | 8.881 a | |

The values with similar letters for each factor or their interactions individually did not differ significantly according to Duncan's test at a probability level of 0.05.

## 4.1.2.4  Fruit moisture percentage (%):

**Table 12** revealed that fig fruit moisture was influenced by fertilizers: Bio Health and Nano Calcium as the application, individual, and interaction. The maximum amount of fruit moisture was 81.528% at the control, without using Bio Health fertilizer, while the lowest amount was 81.119% at the use of 30g tree$^{-1}$ of Bio Health. Furthermore, the maximum amount of fruit moisture was 82.052% at the use of 150mg L$^{-1}$ Nano Calcium, but the lowest value was 80.882% at the control. The interaction effect of 30g tree$^{-1}$ of Bio Health fertilizer and 150mg L$^{-1}$ Nano Calcium recorded the maximum amount of fruit moisture at 82.543%. However, the lowest value was 80.374% at the control.

**Table (12).** Influences of Organic Fertilizer, Nano Calcium, and their interactions on the fruit moisture (%) of figs.

The values with similar letters for each factor or their interactions individually did not differ significantly according to Duncan's test at a  probability level of 0.05.

| Bio Health g tree$^{-1}$ | Nano Calcium mg L$^{-1}$ | | | Average of Bio Health |
|---|---|---|---|---|
| | 0 | 75 | 150 | |
| 0 | 80.374 e | 81.817 bc | 82.393 ab | 81.528 a |
| 15 | 81.897 abc | 80.615 de | 81.220 cd | 81.244 ab |
| 30 | 80.377 e | 80.437 e | 82.543 a | 81.119 b |
| Average of nano calcium | 80.882 b | 80.957 b | 82.052 a | |

## 4.1.2.5   Fruits dropping percentage (%):

The significant effect of Bio Health fertilizer and Nano Calcium on the decrease of fruit dropping is noted in **Table (13).**

The minimum fruit dropping at 30g tree$^{-1}$ of Bio Health fertilizer was 58.364%; however, the maximum dropping recorded at the control was 68.296%. On the other hand, Nano Calcium caused a decrease in fruit dropping; the minimum dropping was 61.978% at the added 150mg L$^{-1}$ of Nano Calcium. The maximum value was 64.346% at the control. The incorporation of both fertilizers also highlighted the significant impacts. The lowest drop recorded at 30g tree$^{-1}$ Bio Health and 150mg L$^{-1}$ Nano Calcium was 56.487%, while the highest drop was 70.017% at the control.

**Table 13:** Influences of Organic Fertilizer, Nano Calcium, and their interactions on the dropping percentage of figs.

| Bio Health g tree$^{-1}$ | Nano Calcium mg L$^{-1}$ | | | Average of Bio Health |
|---|---|---|---|---|
| | **0** | **75** | **150** | |
| **0** | 70.017 a | 68.091 b | 66.780 c | 68.296 a |
| **15** | 62.04 e | 61.497 f | 62.667 d | 62.070 b |
| **30** | 60.981 f | 57.633 g | 56.487 h | 58.364 c |
| **Average of Nano Calcium** | 64.346 a | 62.407 b | 61.978 c | |

The values with similar letters for each factor or their interactions individually did not differ significantly according to Duncan's test at a probability level of 0.05.

### 4.1.2.6 Fruits growth curve depending on diameter(mm$^3$) :

Figure (7) clarified that the interaction between Bio Health fertilizer and Nano Calcium had important effects on fig fruit growth, and they increased in size gradually with time. The combination of 15g tree$^{-1}$ of Bio Health fertilizer and 75g L$^{-1}$ of Nano Calcium determined the largest fruit size from the second to the seventh weeks of the growth duration. However, the lowest fruit size was recorded at the second, third, and fourth weeks of growth in the control without fertilizer. Moreover, the largest fig fruit size was about 57 mm$^3$ at the addition of both fertilizers, 30g tree$^{-1}$ of Bio Health, and 150g L$^{-1}$ of Nano Calcium at the ninth week of fruit growth.

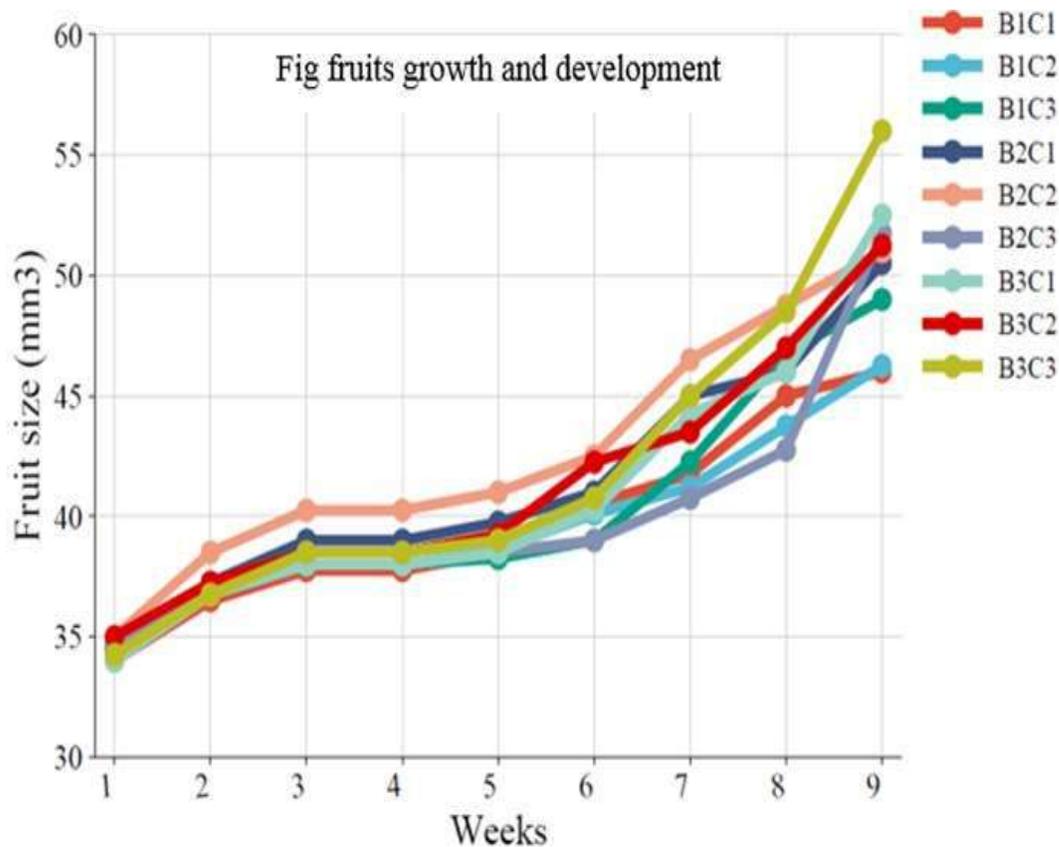

**Figure 7.** Influences on interactions of Organic Fertilizer, Nano Calcium on the growth of fig fruits (mm$^3$).

### 4.1.2.7  Fruit hardness before and after storage (Newton):

**Table (14)** illustrates the effect of Bio Health fertilizer and Nano Calcium individually and in binary on the fruit's hardness before and after storage. The addition of 15g tree$^{-1}$ of Bio Health fertilizer recorded the highest value of fruit hardness before storage, which was 8.612N, and had a significant effect on the control and 30g tree$^{-1}$. However, the highest value of hardness after storage was 9.060N at the control, with a significant influence on both Bio Health treatments. Furthermore, the use of Nano Calcium recorded the highest value of fruit hardness before storage at 8.939 N in the control, with a considerable effect; however, the lowest value was 7.373 N at the addition of 150 mg L$^{-1}$ Nano Calcium. After the fruit was stored, the use of 150mg L$^{-1}$ Nano Calcium showed the highest value of 9.323N and had a significant effect on the control, and 75mg L$^{-1}$ of Nano Calcium.

**Table (14).** Influences of Organic Fertilizer and Nano Calcium on the fruit hardness before and after storage (N) of the figs.

The values with similar letters for each factor or their interactions individually did not differ

| | | Fruit hardness before storage | Fruit hardness after storage |
|---|---|---|---|
| **Bio Health g tree$^{-1}$** | **0** | **8.223 b** | **9.060 a** |
| | **15** | **8.612 a** | **8.338 b** |
| | **30** | **7.658 c** | **8.088 c** |
| **Nano Calcium mg L$^{-1}$** | **0** | **8.993 a** | **7.571 c** |
| | **75** | **8.127 b** | **8.591 b** |
| | **150** | **7.373 c** | **9.323 a** |

significantly according to Duncan's test at a probability level of 0.05.

Considerable effect of both Bio Health fertilizer and Nano Calcium on the fruit hardness before and after storage, at the *P* value level <0.05% **(Figure 8)**

The highest value of fruit hardness before storage was 11.093N at 15g tree$^{-1}$ Bio Health and 0mg L$^{-1}$ Nano Calcium, while the lowest value was 5.920N at 15g tree$^{-1}$ Bio Health and 150mg L$^{-1}$ of Nano Calcium. The highest value of fruit hardness after fruit storage was 10.010N at the 0g tree$^{-1}$ of Bio Health and 150mg L$^{-1}$ of Nano Calcium. However, the lowest value was 6.953N at the 30g tree$^{-1}$ of Bio Health and 0mg L$^{-1}$ Nano Calcium.

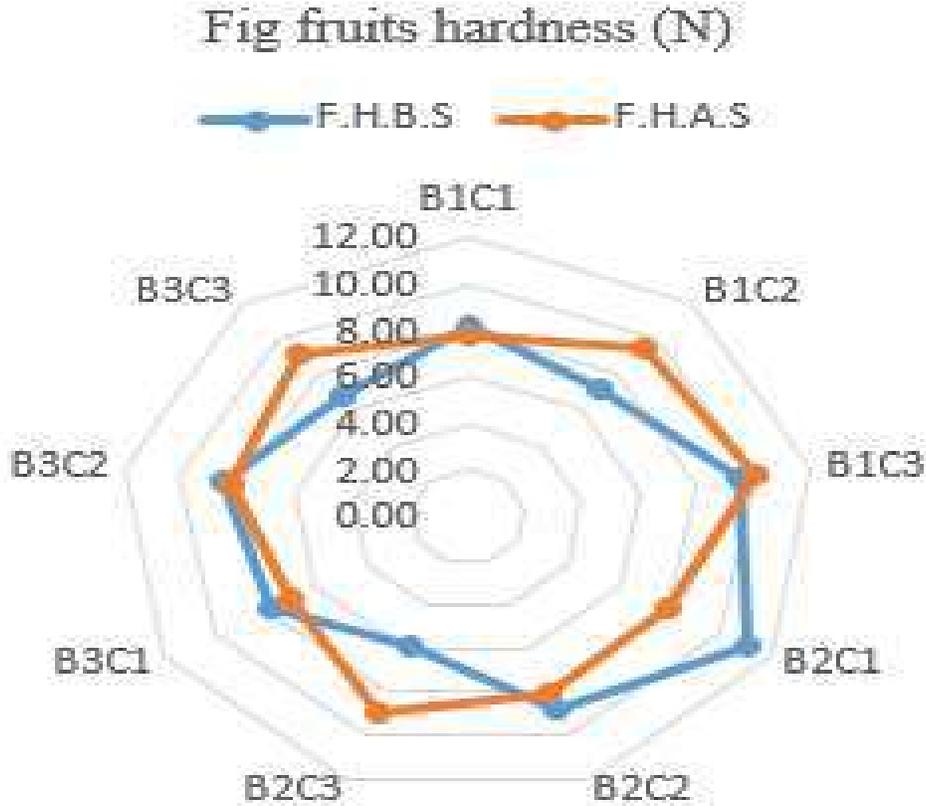

**Figure 8**. Illustrated binary impact of Organic Fertilizer and Nano Calcium on the hardness of the fruit (Newton) before and after storage.

**F.H.B.S**. fruit hardness before storage; **F.H.A.S**. fruit hardness after storage

### 4.1.3 Fig phytochemical characteristics:

#### 4.1.3.1 Impact of Organic Fertilizer on some fruit phytochemical properties Before and after storage:

The significant effect of Bio Health on some of the physicochemical parameters of figs before and after storage, at the level $p$-value <0.05, as seen in **Table 15.** Before fruit storage, the results confirmed that the highest value of TSS was recorded at 30g tree$^{-1,}$ which was 18.33 °Brix, and the lowest was 15.04 °Brix at the control. The highest value of carotenoids was 2.495 µg g$^{-1}$ FW recorded with the use of 15g tree$^{-1}$, but the lowest value was 2.073 µg g$^{-1}$ FW with the use of 30g tree$^{-1}$. On the other hand, the highest value of SSC, 945.194 µg Glu.g$^{-1}$ FW, was given at 15g tree$^{-1}$, but the lowest value was 912.592 µg Glu.g$^{-1}$ FW at the control. The highest value of TPC was 125.544 µg GAE.g$^{-1}$ FW at the control, but the lowest value was 102.410 µg GAE.g$^{-1}$ FW at the use of 15g tree$^{-1}$ of Bio Health fertilizer. Furthermore, the highest amount of AC 224.600 µg TE.g$^{-1}$ FW was recorded by adding 15g tree$^{-1}$, but the lowest amount was at the control was 214.721 µg TE.g$^{-1}$ FW. The main flavonoid found in fig is quercetin; the highest amount, 30.814 µg QE.g$^{-1}$ FW, was shown at the control, but the lowest amount, 25.436 µg QE.g$^{-1}$ FW, was at the 15g tree$^{-1}$. One important antioxidant that is present in figs is ascorbic acid (vitamin C), the highest amount of 3.24 mg 100 g$^{-1}$ FW at the non-treated plant by Bio Health fertilizer (control); however, the lowest amount of 0.532 mg.100 g$^{-1}$ FW at adding 30g tree$^{-1}$. In addition, the highest value of TCC was 23.763g 100 g$^{-1}$ FW at the use of 30g tree$^{-1}$, but the lowest value was 22.963g 100 g$^{-1}$ FW at the adding 15g tree$^{-1}$ of Bio Health fertilizer without showing any significant effects. The highest value of TA at the control was 0.31 (citric acid) %; however, the lowest value was 0.19 (citric acid) % at the use of 30g tree$^{-1}$ of Bio Health fertilizer. Based on the results, the levels of several fruit phytochemicals, such as carotenoids, soluble sugar content (SSC), antioxidant capacity (AC), and total soluble solids (TSS), were considerably raised by Bio Health fertilizer before the fruit was stored.

After fruit storage, the Bio Health fertilizer improved the phytochemical substances in fig fruit significantly, excluding TPC and TFC at the level p-value <0.05 **(Table 15)**

The highest amount of SSC, AC, ASA, carotenoid, TCC, and TA were 690.740 µg Glu.g$^{-1}$ FW, 598.19 µg TE.g$^{-1}$ FW, 29.830 mg g$^{-1}$FW, 1.864 µg g$^{-1}$ FW, 28.637 g g$^{-1}$ FW, and 0.522%(citric acid), respectively, at the use of 30g tree$^{-1}$; however, the lowest of SSC, carotenoid, and TA were 642.733 µg Glu.g$^{-1}$ FW, 1.417 µg g$^{-1}$ FW, and 0.4%, respectively at the control. For AC, ASA, and TCC, the result showed a lower of 515.49 µgTE.g$^{-1}$ FW, 12.921 mg g$^{-1}$FW, and 21.940 g g$^{-1}$ FW, respectively, at the 15g tree$^{-1}$. Furthermore, the control without using Bio Health recorded the highest values of TPC and TFC, which were 186.583 µg GAE.g$^{-1}$ FW and 128.257 µg QE. g$^{-1}$ FW, respectively. But the lowest values were 148.677 µg GAE.g$^{-1}$ FW and 76.903 µg QE. g$^{-1}$ FW, respectively, at the 15g tree$^{-1}$. The highest TSS amount at the use of 15g tree$^{-1}$ was 18.372°Brix, but the lowest was 17.541°Brix at the control. The results indicated a significant increase in the levels of TPC, AC, TFC, TA, and ascorbic acid. However, SSC and carotenoids decreased after fruit storage.

**Table 15**. Influences of Organic Fertilizer on some phytochemical substances

in fig fruit before and after storage

| Before fruit storage | | | | | | | | | |
|---|---|---|---|---|---|---|---|---|---|
| Bio Health g tree⁻¹ | SSC µg g⁻¹ | TPC µg g⁻¹ | AC µg g⁻¹ | TFC µg g⁻¹ | ASA µg g⁻¹ | Carotd µg g⁻¹ | TCC g 100g¹ | TSS% (°Brix) | TA % (Citric acid% %) |
| 0 | 912.592b | 125.544a | 214.721b | 30.814a | 3.24a | 2.233b | 23.751a | 15.04 c | 0.31 a |
| 15 | 945.194a | 102.410c | 224.600a | 25.436b | 0.746b | 2.495a | 22.963a | 16.79 b | 0.24 b |
| 30 | 939.349a | 118.834b | 224.139a | 25.524b | 0.532c | 2.073c | 23.763a | 18.33 a | 0.19 c |
| Means | 932.378 | 115.596 | 221.153 | 27.258 | 1.506 | 2.267 | 23.492 | 16.72 | 0.246 |
| After fruit storage | | | | | | | | | |
| 0 | 642.733b | 186.583a | 557.20b | 128.257a | 14.879a | 1.417b | 22.894b | 17.541b | 0.400c |
| 15 | 678.321a | 148.677c | 515.49c | 76.903c | 12.921c | 1.648ab | 21.940c | 18.372a | 0.465b |
| 30 | 690.740a | 178.157b | 598.19a | 96.853b | 29.830a | 1.864a | 28.637a | 18.357a | 0.522a |
| Means | 690.598 | 171.139 | 556.96 | 100.671 | 19.21 | 1.643 | 24.49 | 18.09 | 0.462 |

The values with similar letters for each factor or their interactions individually did not differ significantly according to Duncan's test at a probability level of 0.05.

SSC, Soluble Sugar Content. TPC, Total Phenol Content. AC, Antioxidant Capacity. TFC, Total Flavonoid Content, ASA, Ascorbic Acid Content. Carotd, Carotenoid Content. TCC Total Carbohydrate Content. TSS, Total Soluble Solid. TA, Total Acidity.

### 4.1.3.2 Influence of Nano Calcium Fertilizer on some fruit phytochemical characteristics before and after fruit storage:

The effect of Nano Calcium on some phytochemical parameters on figs before and after storage was observed as shown in **Table 16** at the p-value <0.05. Before fruit storage, the results confirmed that the highest values of SSC and TPC were (973.880 µg Glu.g⁻¹ FW and 120.248 µg GAE.g⁻¹ FW) were recorded from the control. In addition, the significant value was recorded for (AC and Carotd) after the use of (75mg L⁻¹), which were (246.106 µg TE. g⁻¹ FW and 2.514 µg g⁻¹ FW) respectively, as compared to other parameters. Furthermore, the use of (150mg L⁻¹) showed a significant impact on (ASA,

TCC, and TSS) parameters, which had high rates (1.905 mg g$^{-1}$FW, 24.695g 100g$^{-1}$FW, and 17.45°Brix) respectively, as compared to other parameters. However, no significant value was notified for (TFC) from each treatment. Moreover, after storage samples for 3 weeks, results indicated that some changes were observed (TPC, TFC, and ASA) (186.523 µg GAE.g$^{-1}$ FW, 124.057 µg QE. g$^{-1}$ FW, and 23.850 mg g$^{-1}$) for the control. The highest amount of SSC was measured at 75 mg L$^{-1}$ Nano Calcium was 706.862 µg Glu.g$^{-1}$ FW, compared to the control and 150 mg L$^{-1}$ Nano Calcium. Non-significant value was recorded for carotenoids (1.570, 1.787, and 1.572 µg. g$^{-1}$ FW) among treatments (0, 75, and 150mg L$^{-1}$), respectively. Additionally, no significant rate was also confirmed for TSS as (18.220 and 18.728 °Brix) between (0 and 75mg L$^{-1}$) respectively, while sprayed with (0 and 150mg L$^{-1}$) of Nano Calcium showed no significant rate for TA parameters, which were (0.473 and 0.465%). While the highest value of AC was 582.57 µg QE. g$^{-1}$ FW was recorded at 75mg L$^{-1}$, but the lowest value was 527.72 µg QE. g$^{-1}$ FW at the control. TCC was recorded as the highest value, 26.240g 100g$^{-1}$ FW, after the use of 150mg L$^{-1}$ of Nano Calcium as compared to other parameters. Based on the findings, using nano Calcium increased the levels of several phytochemicals in fig fruit before storage, including AC, TFC, ASA, TCC, TSS, and carotenoid; however, SSC, TPC, and TA decreased. Furthermore, TPC, AC, TFC, ASA, and TA concentrations increased rapidly after storage, but SSC and carotenoid levels generally declined.

**Table 16**: Influence of Nano Calcium spraying on some phytochemical parameters of fig fruit before and after storage.

| Before fruit storage | | | | | | | | | |
|---|---|---|---|---|---|---|---|---|---|
| Nano calcium mg L$^{-1}$ | SSC µg g$^{-1}$ | TPC µg g$^{-1}$ | AC µg g$^{-1}$ | TFC µg g$^{-1}$ | ASA µg g$^{-1}$ | Carotd µg g$^{-1}$ | TCC g 100g$^1$ | TSS% (°Brix) | TA% (Citric acid %) |
| 0 | 973.880a | 120.248a | 213.200b | 27.291a | 1.197c | 2.173b | 21.994a | 15.97c | 0.27a |
| 75 | 907.457b | 116.686b | 246.106a | 26.846a | 1.416b | 2.514a | 23.787a | 16.74b | 0.24b |
| 150 | 915.798b | 109.854c | 204.154c | 27.637a | 1.905a | 2.114c | 24.695a | 17.45a | 0.228c |
| Means | 932.378 | 115.596 | 221.153 | 27.258 | 1.506 | 2.267 | 23.492 | 16.72 | 0.246 |
| After fruit storage | | | | | | | | | |
| 0 | 665.282b | 186.523a | 527.72b | 124.057a | 23.850a | 1.570a | 23.485b | 18.220a | 0.473a |
| 75 | 706.862a | 157.450c | 582.57a | 90.186b | 17.142b | 1.787a | 23.746b | 18.728a | 0.448b |
| 150 | 639.650c | 169.444b | 560.59a | 87.770b | 16.638b | 1.572a | 26.240a | 17.322b | 0.465ab |
| Means | 670.598 | 171.139 | 556.96 | 100.671 | 19.21 | 1.643 | 24.49 | 18.09 | 0.462 |

The values with similar letters for each factor or their interactions individually did not differ significantly according to Duncan's test at a probability level of 0.05.

SSC, Soluble Sugar Content. TPC, Total Phenol Content. AC, Antioxidant Capacity. TFC, Total Flavonoid Content, ASA, Ascorbic Acid Content. Carotd, Carotenoid Content. TCC Total Carbohydrate Content. TSS, Total Soluble Solid. TA, Total Acidity.

## 4.1.3.3 Interaction influences of Organic Fertilizer and Nano calcium on some phytochemical properties of fig fruit before and after storage.

The significant effects illustrated by the Bio Health fertilizer and plant Nano Calcium spraying in fig fruit composition include soluble sugar, total phenolic, and antioxidant capacity before and after the fruit is stored at the ($P > F$) $< 0.05$ **(Table 17).**

The highest values of soluble sugar content (SSC) and total phenol content before fruit storage, 997.453 µg Glu.g$^{-1}$ FW and 160.307 µg GAE.g$^{-1}$ FW, respectively, were recorded by the control. The antioxidant capacity (AC) before fruit storage varied from 172.593 µg TE.g$^{-1}$ FW was recorded by the interaction between 30g tree$^{-1}$ Bio Health and 150mg L$^{-1}$ of Nano Calcium, to 304.317 µg TE.g$^{-1}$ FW, at the interaction between 30g tree$^{-1}$ Bio Health fertilizer and 75mg L$^{-1}$ of Nano Calcium. Furthermore, the highest value of soluble sugar content after the fruit storage, 754.670 µg Glu.g$^{-1}$, was given by the interaction between 30g tree$^{-1}$ Bio Health and 75mg L$^{-1}$ of Nano Calcium. The amount of SSC in the fig fruit dramatically dropped during storage. The highest values of total phenol content and antioxidant capacity, 199.750 µg GAE.g$^{-1}$ and 678.453 µg TE.g$^{-1}$ FW after fruit storage, respectively, were recorded by the interaction between 30g tree$^{-1}$ Bio Health and 150mg L$^{-1}$ of Nano Calcium. The findings showed that the treatment-induced rise in AC differed from the levels of SSC and TPC before fruit storage.

| Before fruit storage |
| --- |

| Bio health g L⁻¹/ nano calcium mg L⁻¹ | SSC µg g⁻¹ | TPC µg g⁻¹ | AC µg g⁻¹ |
|---|---|---|---|
| B1C1 | 997.453 a | 160.307 a | 215.807 d |
| B1C2 | 857.520 d | 97.493 f | 207.393 e |
| B1C3 | 882.80 cd | 118.833 c | 220.963 bcd |
| B2C1 | 950.940 b | 93.933 f | 228.287 b |
| B2C2 | 976.150 ab | 109.640 d | 226.607 bc |
| B2C3 | 908.49 c | 103.657 e | 218.907 cd |
| B3C1 | 973.241 ab | 106.503 de | 195.507 f |
| B3C2 | 888.701 cd | 142.923 b | 304.317 a |
| B3C3 | 956.10 b | 107.073 de | 172.593 g |
| After fruit storage | | | |
| Bio health g L⁻¹/ nano calcium mg L⁻¹ | SSC µg g⁻¹ | TPC µg g⁻¹ | AC µg g⁻¹ |
| B1C1 | 654.501 b | 196.00 a | 505.65 c |
| B1C2 | 684.252 b | 198.500 a | 664.501 a |
| B1C3 | 589.451 c | 165.250 b | 501.450 c |
| B2C1 | 687.502 b | 170.350 b | 512.25 bc |
| B2C2 | 681.671 b | 132.350 d | 532.350 bc |
| B2C3 | 665.80 b | 143.330 c | 501.88 c |
| B3C1 | 653.850 b | 193.220 a | 565.261 b |
| B3C2 | 754.670 a | 141.500 c | 550.850 bc |
| B3C3 | 663.703 b | 199.750 a | 678.453 a |

**Table (17)** The interaction effect of Organic fertilizer and Nano Calcium spraying on fig fruit phytochemical contents (SSC, soluble sugar content, TPC, total phenol content, and AC, antioxidant capacity).
The values with similar letters for each factor or their interactions individually did not differ significantly according to Duncan's test at a probability level of 0.05

SSC, Soluble Sugar Content. TPC, Total Phenol Content. AC, Antioxidant Capacity.

## 4.1.3.4 Combination effects of both Fertilizers on some fig fruit phytochemical traits before and after storage.

The interaction effect of Bio Health fertilizer and plant Nano Calcium spraying on each fig fruit's total flavonoid content, total soluble solids, and total carbohydrate was found to vary considerably. At the $(P > F) < 0.05$ **(Table 18).**

The total flavonoid content (TFC) in fig fruits before storage ranged from 23.037µg QE.g$^{-1}$ FW at the interaction between 15g tree$^{-1}$ Bio Health fertilizer and 75mg L$^{-1}$ of Nano Calcium to 32.44 µg QE.g$^{-1}$FW at the interaction between 0g tree$^{-1}$ Bio Health and 150mg L$^{-1}$ of Nano Calcium. Total soluble solid (TSS) before storage varied between 13.90 °Brix at the control and 19.23 °Brix at the interaction between 30g tree$^{-1}$ Bio Health and 150mg L$^{-1}$ of Nano Calcium, and the total carbohydrate content (TCC) ranged from 20.755 g Glu. 100 g$^{-1}$ FW at the interaction between 15g tree$^{-1}$ Bio Health fertilizer and 0mg L$^{-1}$ of Nano Calcium to 26.107 g Glu. 100 g$^{-1}$ FW before fruit storage at the interaction between 30g tree$^{-1}$ Bio Health fertilizer and 150mg L$^{-1}$ of Nano Calcium. In addition, the highest value of TFC after storage was 162.15 µg QE.g$^{-1}$FW in the control. Still, the lowest value was 62.26 µg QE.g$^{-1}$FW at the interaction between 15g tree$^{-1}$ Bio Health fertilizer and 75mg L$^{-1}$ of Nano Calcium. The total soluble solids after fruit storage ranged between 16.833 °Brix at the interaction between 15g tree$^{-1}$ of Bio Health fertilizer and 150mg L$^{-1}$ of Nano Calcium and 20.433 °Brix at the interaction between 15g tree$^{-1}$ of Bio Health and 75mg L$^{-1}$ of Nano Calcium. The total carbohydrate content before fruit storage differed from 20.755 g 100g$^{-1}$FWat the interaction between 15g tree$^{-1}$ Bio Health and 0mg L$^{-1}$ of Nano Calcium to 26.107g 100g$^{-1}$ FW at the interaction between 15g tree$^{-1}$ Bio Health and 150mg L$^{-1}$ of Nano Calcium; however, TCC after fruit storage ranged between 20.809g 100g$^{-1}$ FW at the interaction between 15 g tree$^{-1}$ Bio Health and 0mg L$^{-1}$ of Nano Calcium and 30.129 g 100g$^{-1}$ FW at the interaction between 30g tree$^{-1}$ Bio Health and 150 mg L$^{-1}$ of Nano Calcium. **Table 18.** The interaction impact of some levels of Organic Fertilizer and Nano Calcium on fig fruit phytochemicals (TFC, total flavonoid, TSS, total soluble solid, and TCC, total carbohydrate content).

| Before fruit storage | | | |
|---|---|---|---|
| Bio health g L$^{-1}$/ nano calcium mg L$^{-1}$ | TFC μg g$^{-1}$ | TSS % (°Brix) | TCC g 100g$^1$ |
| B1C1 | 28.30 cd | 13.90 g | 23.364 b |
| B1C2 | 31.703 ab | 15.23 f | 25.755 a |
| B1C3 | 32.44 a | 16.00 ef | 22.134 c |
| B2C1 | 29.623 bc | 16.43 de | 20.755 c |
| B2C2 | 23.037 g | 16.80 cde | 22.026 c |
| B2C3 | 23.650 fg | 17.13 cd | 26.107 a |
| B3C1 | 23.950 fg | 17.57 bc | 21.864 bc |
| B3C2 | 25.80 ef | 18.20 b | 23.850 b |
| B3C3 | 26.820 de | 19.23 a | 25.845 a |
| After fruit storage | | | |
| Bio health g L$^{-1}$/ nano calcium mg L$^{-1}$ | TFC μg g$^{-1}$ | TSS% (°Brix) | TCC g 100g$^1$ |
| B1C1 | 162.15 a | 18.033 bc | 22.13 |
| B1C2 | 113.070 b | 17.11 c | 23.04 d |
| B1C3 | 112.550 b | 17.480 bc | 25.37 c |
| B2C1 | 99.800 c | 17.850 bc | 20.809 |
| B2C2 | 62.260 f | 20.433 a | 27.88 b |
| B2C3 | 68.650 e | 16.833 c | 27.09 b |
| B3C1 | 113.220 b | 18.776 b | 24.31 c |
| B3C2 | 95.230 c | 18.614 b | 27.55 b |
| B3C3 | 82.110 d | 17.667 bc | 30.129 a |

The values with similar letters for each factor or their interactions individually did not differ significantly according to Duncan's test at a probability level of 0.05.

TFC, Total Flavonoid Content; TSS, Total Soluble Solid; TCC, Total Carbohydrate Content.

## 4.1.3.5 Binary impact of Organic Fertilizer and Nano Calcium on some fruit phytochemical traits at both conditions.

A considerable variation was observed among the effects of different levels of Bio Health fertilizer and Nano Calcium spraying applications on fig fruit's ascorbic acid, carotenoid content, and titratable acidity, at the level of (P > F) < 0.05 **(Table 19).**

The ascorbic acid content in fig fruits varied between 0.207 mg.100g$^{-1}$ FW was recorded at the interaction between 30g tree$^{-1}$ Bio Health fertilizer and 75mg L$^{-1}$ of Nano Calcium, and 4.35 mg 100g$^{-1}$ at the interaction between 0g tree$^{-1}$ Bio Health fertilizer and 150mg L$^{-1}$ of Nano Calcium before storage. Still, after storing fruits, the ascorbic acid significantly increased, ranging between 9.57 mg/100g$^{-1}$ FW at the control and 49.55mg.100g$^{-1}$ FW at the interaction between 30g tree$^{-1}$ Bio Health fertilizer and 0mg L$^{-1}$ of Nano Calcium. However, the fruit carotenoid content before storage varied from 1.857 µg g$^{-1}$ FW at the interaction between 30g tree$^{-1}$ Bio Health fertilizer and 150mg L$^{-1}$ of Nano Calcium to 2.717 µg g$^{-1}$ FW at the interaction between 15g tree$^{-1}$ Bio Health fertilizer and 75mg L$^{-1}$ of Nano Calcium. However, after storage, it ranged between 0.980 µg g$^{-1}$ FW at the control. and 1.97 µg g$^{-1}$ FW at the interaction between 30g tree$^{-1}$ Bio Health fertilizer and 75g L$^{-1}$ of Nano Calcium. Furthermore, the lowest value of fruit titratable acidity (TA) before storage, 0.174%, was recorded by the interaction of 30g tree$^{-1}$ Bio Health fertilizer and 150mg L$^{-1}$ of Nano Calcium. But the highest value was 0.35% at the control. The results indicated that using Bio Health fertilizer and Nano Calcium gradually decreased TA. After fruit storage, the highest value of 0.553% was given by the interaction of 30g tree$^{-1}$ Bio Health fertilizer and 0mg L$^{-1}$ of Nano Calcium; however, the lowest value of 0.36% was given by the 0g tree$^{-1}$ Bio Health fertilizer and 150mg L$^{-1}$ of Nano Calcium.

**Table 19.** The interaction effects of Organic Fertilizer and Nano Calcium spraying on fig fruit phytochemicals (ASA, ascorbic acid, Carotd, carotenoids, and TA, titratable acid content).

| Before fruit storage | | | |
|---|---|---|---|
| **Bio health g L$^{-1}$/ nano calcium mg L$^{-1}$** | **ASA** <br> µg g$^{-1}$ | **Carotd** <br> µg g$^{-1}$ | **TA%** <br> (Citric acid %) |
| **B1C1** | 2.293 c | 2.213 d | 0.35 a |
| **B1C2** | 3.076 b | 2.383 c | 0.30 b |
| **B1C3** | 4.35 a | 2.103 e | 0.27 c |
| **B2C1** | 0.346 g | 2.386 c | 0.25 cd |
| **B2C2** | 0.963 d | 2.717 a | 0.24 d |
| **B2C3** | 0.930 e | 2.389 c | 0.241 d |
| **B3C1** | 0.953 de | 1.920 f | 0.21 e |
| **B3C2** | 0.207 h | 2.443 b | 0.19 f |
| **B3C3** | 0.436 f | 1.857 g | 0.174 g |
| After fruit storage | | | |
| **Bio health g L$^{-1}$/ nano calcium mg L$^{-1}$** | **ASA** <br> µg g$^{-1}$ | **Carotd** <br> µg g$^{-1}$ | **TA%** <br> (Citric acid %) |
| **B1C1** | 9.57 f | 0.980 c | 0.39 cd |
| **B1C2** | 16.377 de | 1.886 a | 0.45 c |
| **B1C3** | 18.290 cd | 1.270 bc | 0.36 d |
| **B2C1** | 12.027 f | 1.820 a | 0.486 c |
| **B2C2** | 15.830 e | 1.510 abc | 0.381 d |
| **B2C3** | 10.907 f | 1.616 ab | 0.530 ab |
| **B3C1** | 49.55 a | 1.793 a | 0.553 a |
| **B3C2** | 19.22 bc | 1.970 a | 0.516 b |
| **B3C3** | 20.720 b | 1.830 a | 0.497 b |

The values with similar letters for each factor or their interactions individually did not differ significantly according to Duncan's test at a probability level of 0.05

ASA, Ascorbic Acid Content; Carotd, Carotenoid Content; TA, Total Acidity

**4.1.3.6 The Ratio of total soluble solids (TSS) / total acidity (TA) of fruit juice before storage**.

The results in **Table (20)** observed the significant influence of Bio Health fertilizer on the TSS /TA ratio. The fertilizer treatment (0g tree$^{-1}$) had the highest ratio (80.309, 72.843) for both Bio Health fertilizer (0g tree$^{-1}$) and Nano Calcium (0mg L$^{-1}$), while low values were obtained from concentrations (30g tree$^{-1}$) of Bio Health and (150mg L$^{-1}$) of Nano Calcium which had the lowest value (60.159 and 66.033) respectively.  In terms of the interaction between Bio Health fertilizer and Nano Calcium, the results showed that no significant effect was demonstrated from the interaction between (0g tree$^{-1}$) Bio Health and (0mg L$^{-1}$) of Nano Calcium with (0g tree$^{-1}$) Bio Health and (75 mg L$^{-1}$) of Nano Calcium which were verified the highest TSS/TA ratio (84.148 and 81.722) respectively as compared to other interaction treatments.

**Table 20.** Influences of Organic Fertilizer, Nano Calcium, and their interactions on the ratio of TSS/TA of fig fruits**.**

| Bio Health g tree$^{-1}$ | Nano Calcium mg L$^{-1}$ | | | Average of Bio Health |
|---|---|---|---|---|
| | **0** | **75** | **150** | |
| **0** | 84.148 a | 81.722 a | 75.057 b | 80.309 a |
| **15** | 68.472 cd | 70.000 c | 68.411 cd | 68.961 b |
| **30** | 65.909 d | 59.938 e | 54.630 f | 60.159 c |
| **Average of Nano Calcium** | 72.843 a | 70.554 b | 66.033 c | |

The values with similar letters for each factor or their interactions individually did not differ significantly according to Duncan's test at a  probability level of 0.05.

**4.1.4 Storage qualities:**

### 4.1.4.1 Percentage of fruit weight loss and non-suitable fruit for marketing:

Important impacts of both fertilizers were demonstrated in the decrease of fruit weight loss and fruit non-suitable for markets in **Table (21).**

The minimum amount of fruit weight loss and non-useful for marketing were 11.490% and 17.778%, respectively, at the addition of 30g of Bio Health fertilizer per tree, while the maximum amounts at the control were 18.644% and 22.732%, respectively. In addition, the addition of 150mg $L^{-1}$ of Nano Calcium gave the minimum value of fruit weight loss, 13.848%, but the maximum values of fruit weight loss and fruit non-marketable were 15.770% and 24.487%, respectively, at the control. However, the minimum value of non-marketable fruit was 14.178% at the use of 75mg $L^{-1}$ of Nano Calcium.

**Table (21).** Influences of Organic Fertilizer, Nano Calcium on the decreased amount of fruit loss and non-marketable percentage.

|  |  | % weight loss Fruits | % non-marketable Fruit |
|---|---|---|---|
| **Bio Health g tree$^{-1}$** | **0** | **18.644 a** | **22.732 a** |
|  | 15 | **14.229 b** | **18.503 b** |
|  | 30 | **11.490 c** | **17.778 b** |
| **Nano Calcium mg L$^{-1}$** | **0** | **15.770 a** | **24.487 a** |
|  | 75 | **14.746 b** | **14.178 c** |
|  | 150 | **13.848 c** | **20.349 b** |

The values with similar letters for each factor or their interactions individually did not differ significantly according to Duncan's test at a probability level of 0.05.

Significant effects of Bio Health fertilizer and Nano Calcium on reducing fruit loss weight (%) and non-suitable fruits for marketing (%), at the *P* value level <0.05% **(Figure 9).**

The minimum amount of weight loss was 11.170% of fruit weight obtained at the 30g tree$^{-1}$ of Bio Health fertilizer and 150mg L$^{-1}$ Nano Calcium. However, the maximum amount was 20.237% of fruit weight at the control. In addition, the lowest amount of fruits Non-Suitable for marketing was 11.183% of fruits stored recorded at the 15g tree$^{-1}$ of Bio Health fertilizer and 75mg L$^{-1}$ of Nano Calcium, while the highest amount of fruits not suitable for marketing was 27.673% of storage fruits at the control.

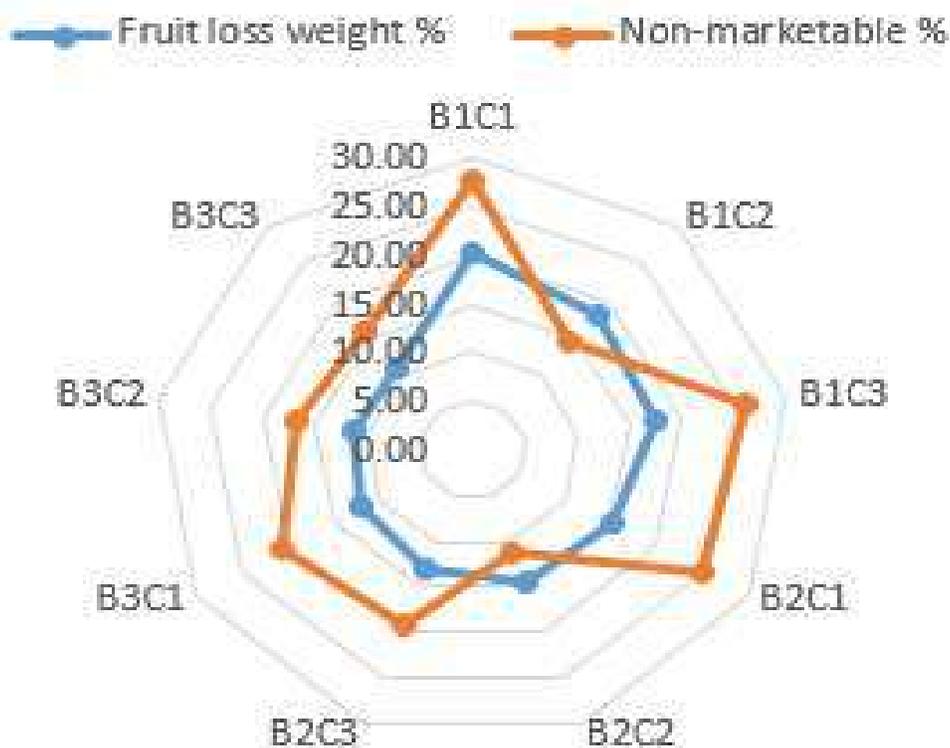

**Figure 9.** Combined effects of Organic Fertilizer and Nano calcium on the percentage of fruit loss weight and non-marketable fruits.

## 4.2 Discussion

This study's focus enhanced the qualities and quantities of a particular fig cultivar by applying Bio Health fertilizer and spraying the plants with Nano Calcium. Nano fertilizers have supported increased productivity by allowing targeted nutrient delivery, ensuring a gradual release, and improving nutrient efficiency while reducing overall fertilizer usage. (Verma *et al*., 2022).

Biofertilizers improve plant growth through direct mechanisms like nitrogen fixation, phytohormone production, and phosphate solubilization, while indirect mechanisms protect plants from pathogens via antibiotics, HCN, siderophores, and lytic enzymes (Mahmud *et al*., 2021). Biofertilizers promote plant growth and enhance crop yield by improving nitrogen availability and producing growth-stimulating substances such as auxins, cytokinins, and gibberellins (Basalingappa *et al*., 2018).

The study's findings showed that the application of organic fertilizer (Bio Health) enhanced both vegetative traits (Leaf area, total chlorophyll content, and leaf dry weight) and nutrient quantity characteristics (fruit weight, fruit volume, and total yield) significantly. This improvement is credited to the fertilizer's important macronutrients (potassium, phosphorus, and nitrogen), micronutrients, and organic acids (fulvic and humic acids), which are critical for enhancing microbial activity and soil fertility. These elements increase the nutrients' availability to the plant by facilitating their progressive release. Nitrogen availability in particular stimulated the production of proteins and chlorophyll, which increased photosynthetic efficiency and had a beneficial effect on leaf area and dry weight. The development of roots and the stimulation of essential metabolic activities, particularly cell division, were facilitated by phosphorus, whereas potassium was essential for controlling physiological processes and delivering photosynthetic products to the fruits, which resulted in increased fruit weight and volume. According to earlier studies, organic fertilizers have a positive impact on plant growth and production by improving soil health and nutrient cycling (Zhang *et al*., 2016; Adesemoye *et al*., 2009). These findings are consistent with those findings.

Agrochemicals known as "Nano fertilizers" are made of particles between 100 and 250 nm in size. They improve long-term results, dissolve more readily in water, and reduce element loss through leaching and emissions (Al-Rekaby and Atiyah, 2020).

The Bio Health in this study contains specific strains of Trichoderma and Bacillus, both of which play vital roles in improving soil fertility; Trichoderma is known for its ability to decompose organic matter, suppress plant diseases, and improve plant growth by promoting nutrient availability. (Kamal, 2018).

Bacillus strains contribute to soil health by fixing nitrogen, producing beneficial enzymes, and enhancing plants' uptake of nutrients; Together, these microorganisms generate a synergistic effect that improves soil quality and supports overall plant health and productivity. (Santoyo *et al*., 2024).

Enzymes like amylases and invertases, which break down complex carbs into simpler sugars that can be used for respiration or other metabolic functions, become more active after fruit is stored. Under stress, carbohydrates can also be broken down by oxidative enzymes. (Rasool *et al.*, 2023).

In physiological and Biochemical processes, Nano fertilizers are important because they increase nutrient availability, improving metabolic processes and boosting meristematic activities, increasing the photosynthetic area and apical development, It is essential for boosting flowering, vegetative development, and reproductive growth, as well as improving fruit yield, product quality, and shelf life. (Shilpa, *et al.*, 2022; Hakeem, *et al.*, 2023).

Figs treated with $Ca^{+2}$ exhibit significant levels of antioxidant activity because $Ca^{+2}$ causes reactive oxygen species to be produced by activating NADPH oxidase. As of the defense mechanisms triggered by the resultant oxidative stress, fig health is improved by minimizing oxidative damage and enhancing phenolic production and antioxidant enzyme activity (Souza *et al*., 2023). Which returns to the role of ascorbic acid and acts as a key antioxidant by maintaining active forms of other antioxidants. However, it easily degrades during storage and handling. Proper postharvest practices are essential to retain their levels in fruits and vegetables, preserving their nutritional value (Badgujar *et al.*, 2014).

Due to starch breakdown into sugars and water loss through transpiration. These changes concentrate sugars, illustrating how storage impacts fruit quality, particularly in figs. (Irfan *et al.*, 2013). Phenylalanine ammonia-lyase (PAL), a crucial enzyme in the phenylpropanoid pathway that produces flavonoids, is regulated by Calcium through the activation of calmodulin and other calcium-binding proteins. (Zheng *et al.*, 2022). Calcium can reduce the severity of fruit respiration and enhance fruit storage quality by weakening the metabolism of endogenous chemicals in fruits and increasing their oxidative capacity. According to this study, the results of using Nano Calcium were better than those of the controls. Additionally, calcium controls enzyme activity and the ionic environment, both of which impact how well fruits store (Petkova *et al.*, 2019; Zhang *et al.*, 2019a). Ascorbic acid, vitamin C, is a main antioxidant broadly found in plants. It has been shown that the mitochondria produce ascorbic acid through many hypothesized pathways. The primary biosynthetic pathway is the L-galactose pathway, commonly known as the Smirnoff–Wheeler pathway. (Steven Hill *et al.*, 1998; Zheng *et al.*, 2022). During fruit storage, the enzyme L-galactono-1,4-lactone dehydrogenase (GLDH) increases ascorbic acid synthesis, improving nutritional quality. Additionally, sugar breakdown during ripening further boosts ascorbic acid levels,  improving flavor and health benefits. (Zheng *et al.*, 2022).

Figs include carotenoids such as lutein, zeaxanthin, ß-cryptoxanthin, and ß-carotene. (Yemiş *et al.*,  2012; Rasool *et al.*, 2023). They discovered that these pigments are present in yellow fig cultivars and that the ripening stage affects the fig fruits' surface color. In addition, the results indicated that using bio-health fertilizer and nano calcium gradually decreased TA. After fruit storage, the data

also showed that the Nano calcium caused a decrease in TA in fruit, but after fruit storage, TA slightly increased. Our results were similar to those reached by (Irfan *et al*., 2013; Hssaini *et al*., 2019), but (Zhang *et al*., 2019a) showed that the TA decreased after fruit storage. Enzymatic activity and environmental elements, including temperature, light, and nutrient balance, control titratable acidity. This study indicated that ascorbic acid increased after fruit storage. Titratable acidity in fruits is largely influenced by the synthesis and storage of organic acids, such as citric and malic acids, in vacuoles. *(*Walia *et al.*, 2022). Our study found that using a combination of organic fertilizer (Bio Health) and nano-calcium foliar spray greatly improved many of the key parameters we looked at this including leaf area, chlorophyll content, and leaf dry weight, as well as fruit weight, volume, total yield, and postharvest qualities like fruit firmness. The combination also reduced fruit weight loss and spoilage during storage. This combined approach consistently outperformed either treatment on its own, showing a synergistic effect. These improvements can be credited to the combined physiological and nutritional benefits of both inputs. The Bio Health organic fertilizer helps create a better soil structure, boosts microbial activity, and releases nutrients gradually, all of which support long-term plant growth and fruit development (Chen, 2006; Adesemoye *et al*., 2009). At the same time, applying nano-calcium plays a crucial role in improving nutrient absorption and maintaining cell membrane health, while also strengthening cell walls, which enhances fruit quality and storage potential (White and Broadley, 2003; Riaz *et al*., 2020).These results are consistent with previous studies suggesting that integrating organic fertilizers with nano-fertilizers can improve plant performance, productivity, and fruit quality (Saber *et al*., 2012; Hassan *et al*., 2021). In this study, the combined treatment recorded the highest values across most measured traits, emphasizing its effectiveness in supporting optimal growth and productivity of *Ficus carica* L. (Wazeri cultivar) grown under greenhouse conditions.

**Conclusion**

This investigation clarified that the individual and combined application of organic fertilizer (Bio Health) and plants sprayed with Nano Calcium significantly enhanced the growth, yield, quality, and phytochemical content of figs under plastic house conditions.

- **Vegetative growth:** Both treatments, organic fertilizer (Bio Health) and Nano Calcium, in individual and binary shapes, increased leaf area, chlorophyll content, leaf dry weight, and nutrient uptake (nitrogen, phosphorus, and potassium), contributing to overall plant health.

- **Fruit quality:** Both treatments enhanced fruit weight, volume, and moisture content, while reducing fruit drop and weight loss, increasing suitable marketable yield.

- **Phytochemical properties**: The fig fruits exhibited improved antioxidant capacity, total flavonoid content, and ascorbic acid levels by application of the mentioned treatments, reduction of Total Acidity in fruits. As well as increasing their nutritional and health benefits.

- **Post-harvest improvements:** Both treatments effectively conserved fruit quality during storage, especially Nano Calcium, reducing deterioration and prolonging shelf life.

- **Optimal application**: The best outcomes were found at the interaction between 30 g tree$^{-1}$ of Bio Health fertilizer, 150 mg L$^{-1}$ of Nano Calcium, and 75 mgL$^{-1}$ Nano Calcium alone, maximizing fruit growth, phytochemical content, and storage stability. The commonly obtained higher leaf chlorophyll content due to the fertilizers resulted in increased photosynthesis, leading to better vegetative growth and fruit development. These discoveries highlight the importance of single and incorporating Bio Health fertilizer and Nano fertilizer applications in modern fig agribusiness for sustainable and high-quality production.

**Recommendations**

- Adding 30 g tree$^{-1}$ of Bio Health fertilizer and 150 mg L$^{-1}$ of Nano Calcium as binary and 75mg L$^{-1}$ individually for maximum growth, fruit quality, and post-harvest longevity.
- Advised to carry out this experiment in other places, especially those where winter is not cold waves are postponed, by using different doses of both Organic Fertilizer and Nano Calcium.
- Maintain constant watering calendars and pruning practices to support optimal nutrient absorption, decrease fruit drop, and enhance yield quality.
- Incorporate Bio Health fertilizers and Nano fertilizers with environmentally friendly techniques to improve soil fertility, reduce chemical inputs, and improve fig production efficiency.
- Conduct more research in the future to discover the effect of Bio Health fertilizer and Nano Calcium on the vegetative, quality, and phytochemical properties of figs.

——————— **REFERENCES** ———————

Mission. Journal of Agriculture and Ecology Research International, 23(6), 13–23. https://doi.org/10.9734/JAERI/2022/v23i6505

Najim, M. H., Khalaf, R. A., and Al-Zubaidi, K. A. (2022). Response of fig trees to foliar application of potassium humate under field conditions. Diyala Agricultural Sciences Journal, 14(1), 88–97.

Oliveira, M., de Souza, D. T. V., dos Santos, A. F., and Rodrigues, A. S. (2023). Calcium nutrition in fig orchards enhances fruit quality at harvest and storage. Plants, 9(1), 123. https://doi.org/10.3390/plants9010123

Osman, S. M., and Abd El-Rhman, I. E. (2010). Effect of organic and bio-N-fertilization on growth and productivity of fig tree (Ficus carica L.). Research Journal of Agriculture and Biological Sciences, 6(3), 3195–328.

Otieno, D. F., and Mwangi, G. M. (2023). Effect of foliar application of calcium chloride on vegetative growth, chlorophyll content, and nitrogen status in fig trees (Ficus carica L.). Journal of Horticultural Sciences, 18(2), 145–153. https://doi.org/10.1007/s42789-023-00215-7

Özdemir, E., and Bayındırlı, A. (2018). Effect of pre-harvest humic acid applications on postharvest quality and shelf life of fresh fig (Ficus carica L.). Scientia Horticulturae, 234, 234–240.

Özdemir, E., and Bayındırlı, A. (2025). Impact of foliar application of calcium and boron on growth, nutrient content, and fruit quality of "Sultani" figs under saline conditions. Agricultural Science and Soil Sciences, 13(1), 752–762.

Pauly, N., Schmidt, R., Mariusz, P., and Martinoia, E. (2000). Control of free calcium in plant cell nuclei. Nature, 405(6788), 754–755. https://doi.org/10.1038/35015671

Pereira, C., Ribeiro, L., and Sousa, M. (2017). Influence of ripening stage on bioactive compounds and antioxidant activity in nine figs (Ficus carica L.) varieties grown in Extremadura, Spain. Journal of Food Composition and Analysis, 64, 203–212. https://doi.org/10.1016/j.jfca.2017.09.006

Petkova, N., Ivanov, I., and Denev, P. (2019). Changes in phytochemical compounds and antioxidant potential of fresh, frozen, and processed figs (Ficus carica L.). International Food Research Journal, 26(2), 1881–1888. doi: 10.15722/jfsci.26.2.20190201

──────── **REFERENCES** ────────

L.). Journal of Agricultural and Food Chemistry, 54(20), 7717–7723. https://doi.org/10.1021/jf060497h

**Souza, J.M.A., Leonel, S., Leonel, M., Garcia, E.L., Ribeiro, L.R., Ferreira, R.B., Martins, R.C., de Souza Silva, M., Monteiro, L.N.H. and Duarte, A.S., 2023**. Calcium nutrition in fig orchards enhances fruit quality at harvest and storage. Horticulture, 9(1), 123. https://doi.org/10.3390/horticulturae9010123

**Spinelli, F.; F. Giovanni; N. Massimo; S. Mattia and C. Guglielmo. (2009).** Perspectives on the use of a seaweed extract to moderate the negative effects of alternate bearing in apple trees. Journal of Horticultural Science and Biotechnology, Special Issue, 17(1), 131-137.

**Steven Hill, S. S., Sweetlove, L., and Clair, S. G. (1998).** Physiology and metabolism: Engineering secondary metabolism in maize cells by ectopic expression of transcription factors. Strawberry Cultivar Festival. Master's thesis, College of Agriculture, University of Baghdad, Iraq.

**Taha and S. M. (2008).** The effect of spraying with gibberellic acid, cystocele, and three extracts from marine plants on some vegetative and flowering growth characteristics and yield components of two strawberry (*Fragaria × ananassa* Duch.) cultivars. Ph.D. Dissertation, Department of Horticulture, College of Agriculture, Salahaddin University.

**Taher and Hassan. (2005).** Effect of different doses of sources of boric acid on growth and yield. Agricultural Science Digest, 28(3), 189-191.

**Taher, A. W., and Ethbeab, I. J. (2021).** The effect of the bio-fertilizer Seek on some vegetative characteristics of seedlings of two fig cultivars (*Ficus carica* L.). International Journal of Agricultural and Statistical Sciences, 17(1), 2163–2168.

**Tahir, N. A. R., Rasul, K. S., Lateef, D. D., and Grundler, F. M. (2022)**. Effects of oak leaf extract, biofertilizer, and soil containing oak leaf powder on tomato growth and biochemical characteristics under water stress conditions. Agriculture, 12(12), 2082. https://doi.org/10.3390/agriculture12122082

**Tahir, N. A. R., Rasul, K. S., Lateef, D. D., Aziz, R. R., and Ahmed, J. O. (2024).** In vitro evaluation of Iraqi Kurdistan tomato accessions under drought stress conditions using polyethylene glycol-6000. Life, 14(11), 1502. https://doi.org/10.3390/life14111502

**Taiz, L. and Zeiger, E. (2006).** Plant physiology. 4thed. Sinauer Associates, Inc., publishers, Sunderland, Massachusetts.